\documentclass[twocolumn]{aastex631}

\usepackage{xcolor}

\usepackage{subfigure}

\usepackage{multirow}
\shorttitle{AASTeX v6.31 Sample article}
\shortauthors{Xie et al.}
\newcommand{\uat}[2]{\href{http://vocabs.ands.org.au/repository/api/lda/aas/the-unified-astronomy-thesaurus/current/resource.html?uri=http://astrothesaurus.org/uat/#1}{#2  (#1)}}
\usepackage[T1]{fontenc}
\graphicspath{{./}{figures/}}

\usepackage[most]{tcolorbox}

\newtcbox{\myboxi}[1][]{nobeforeafter,tcbox raise 
base,colframe=black!50!black,colback=black!50!black,height=8pt,valign=center,raster 
valign=center,
box align=base,sharp corners,top=0pt,bottom=0pt,left=0pt,right=2pt,
boxrule=0pt,boxsep=2.5pt,before upper=\strut,#1}
\newcommand{\myboxblk}[2][1.1ex]{\raisebox{#1}{\myboxi{#2}}}

\newtcbox{\myboxxi}[1][]{nobeforeafter,tcbox raise 
base,colframe=black!50!black,colback=black!40!white,height=8pt,valign=center,raster 
valign=center,
box align=base,sharp corners,top=0pt,bottom=0pt,left=0pt,right=2pt,
boxrule=0pt,boxsep=2.5pt,before upper=\strut,#1}
\newcommand{\myboxgray}[2][1.1ex]{\raisebox{#1}{\myboxxi{#2}}}

\newtcbox{\myboxxxi}[1][]{nobeforeafter,tcbox raise 
base,colframe=black!85!black!,colback=white!85!white!,height=8pt,valign=center,raster 
valign=center,
box align=base,sharp corners,top=0pt,bottom=0pt,left=0pt,right=2pt,
boxrule=0pt,boxsep=2.5pt,before upper=\strut,#1}
\newcommand{\myboxwhi}[2][1.1ex]{\raisebox{#1}{\myboxxxi{#2}}}

\begin{document}

\title{From Young to Older Disks: JWST/MIRI Evidence for Fading Molecular Emission and Hints for Elevated C/O in Upper Scorpius}

\newcommand{\affilLPL}{\affiliation{Lunar and Planetary Laboratory, The University of Arizona, Tucson, AZ 85721, USA; \url{cyxie@arizona.edu}}}

\correspondingauthor{Chengyan Xie}
\email{cyxie@arizona.edu}

\author[0000-0001-8184-5547]{Chengyan Xie}
\affilLPL

\author[0000-0001-7962-1683]{Ilaria Pascucci}
\affilLPL

\author[0000-0002-7607-719X]{Feng Long}
\affilLPL
\affiliation{Kavli Institute for Astronomy and Astrophysics, Peking University, Beijing 100871, China}

\author{Uma Gorti}
\affiliation{NASA Ames Research Center, Moffett Field, CA 94035, USA}
\affiliation{Carl Sagan Center, SETI Institute, Mountain View, CA 94043, USA}

\author[0000-0003-4335-0900]{Andrea Banzatti}
\affiliation{Department of Physics, Texas State University, 749 North Comanche Street, San Marcos, TX 78666, USA}

\author[0000-0002-0364-937X]{Richard Booth}
\affiliation{School of Physics and Astronomy, University of Leeds, Leeds, LS2 9JT, UK}

\author[0000-0001-7552-1562]{Klaus M. Pontoppidan}
\affiliation{Jet Propulsion Laboratory, California Institute of Technology, 4800 Oak Grove Drive, Pasadena, CA, 91109, USA}

\author[0000-0003-0448-6354]{Tamara Molyarova}
\affiliation{School of Physics and Astronomy, University of Leeds, Leeds, LS2 9JT, UK}

\author[0000-0003-2251-0602]{John Carpenter}
\affiliation{Joint ALMA Observatory, Avenida Alonso de Córdova 3107, Vitacura, Santiago, Chile}

\author[0000-0001-8060-1321]{Min Fang}
\affiliation{Purple Mountain Observatory, Chinese Academy of Sciences, 10 Yuanhua Road, Nanjing 210023, People’s Republic of China}
\affiliation{University of Science and Technology of China, Hefei 230026, People’s Republic of China}

\author[0000-0002-7616-666X]{Yao Liu}
\affiliation{School of Physical Science and Technology, Southwest Jiaotong University, Chengdu 610031, China}

\author[0009-0002-2380-6683]{Eshan Raul}
\affiliation{Department of Astronomy, University of Wisconsin--Madison, Madison, WI 53706, USA}

\author[0000-0002-0661-7517]{Ke Zhang}
\affiliation{Department of Astronomy, University of Wisconsin--Madison, Madison, WI 53706, USA}

\author[0000-0002-2314-7289]{Steve Ertel}
\affiliation{Department of Astronomy and Steward Observatory, University of Arizona, 933 N Cherry Ave., Tucson, AZ 85721-0065, USA}
\affiliation{Large Binocular Telescope Observatory, University of Arizona, 933 N Cherry Ave., Tucson, AZ 85721-0065, USA}

\author[0000-0003-0454-3718]{Jordan Stone}
\affiliation{Naval Research Laboratory, Remote Sensing Division, 4555 Overlook Ave SW, Washington, DC 20375, USA}
\affiliation{Department of Astronomy and Steward Observatory, University of Arizona, 933 N. Cherry Ave, Tucson, AZ 85721-0065, USA}

\author[0000-0001-6448-7178]{Aaron Empey}
\affiliation{University College Dublin (UCD), Department of Physics, Belfield, Dublin4, Ireland}

\author[0000-0003-3562-262X]{Carlo F. Manara}
\affiliation{European Southern Observatory, Karl-Schwarzschild-Strasse 2, 85748 Garching bei München,Germany}

\author[0000-0001-8764-1780]{Paola Pinilla}
\affiliation{Mullard Space Science Laboratory, University College London, Holmbury St Mary, Dorking, Surrey RH5 6NT, UK}

\author[0000-0003-3682-6632]{Colette Salyk}
\affiliation{Department of Physics and Astronomy, Vassar College, 124 Raymond Avenue, Poughkeepsie, NY 12604, USA}

\author[0000-0002-1103-3225]{Benoit Tabone}
\affiliation{Université Paris-Saclay, CNRS, Institut d’Astrophysique Spatiale, 91405 Orsay, France}

\author[0000-0002-4147-3846]{Miguel Vioque}
\affiliation{European Southern Observatory, Karl-Schwarzschild-Strasse 2, D-85748 Garching bei München, Germany}

\author[0000-0002-2828-1153]{Lucas Cieza}
\affiliation{Instituto de Estudios Astrofísicos, Universidad Diego Portales, Av. Ejercito 441, Santiago, Chile}

\author[0000-0003-4853-5736]{Giovanni Rosotti}
\affiliation{Dipartimento di Fisica, Università degli Studi di Milano, Via Celoria 16, I-20133 Milano, Italy}

\author[0000-0002-1575-680X]{James Miley}
\affiliation{Joint ALMA Observatory, Alonso de Córdova, 3107, Vitacura, Santiago, Chile}
\affiliation{European Southern Observatory, Alonso de Córdova, 3107, Vitacura, Santiago, Chile
Millennium Nucleus on Young Exoplanets and their Moons (YEMS), Chile}

\author[0000-0003-0787-1610]{Geoffrey A. Blake}
\affiliation{Division of Geological and Planetary Sciences, California Institute of Technology, MC 150-21, 1200 East California Boulevard, Pasadena, CA 91125, USA}

\author[0000-0002-1566-389X]{Abygail Waggoner}
\affiliation{Department of Astronomy, University of Wisconsin--Madison, Madison, WI 53706, USA}

\begin{abstract}
We present JWST/MIRI spectroscopy of 14 disks in the older ($\sim$5–10 Myr) Upper Scorpius (USco) association and  use slab  of gas in local thermal equilibrium to infer basic gas properties. We find that half of these disks are molecular rich, with detections of H$_2$O, CO$_2$, HCN, C$_2$H$_2$, and H$_2$, while the other half are molecular poor, showing no molecular emission other than H$_2$. 
We further combine this sample with 10 other USco disks from the AGE-PRO program and compare the combined older sample to young ($\sim 1-3$\,Myr) JDISCS Cycle~1 systems, which are analyzed in a similar manner. 
We find that USco disks have lower detection rates of major molecular species but a significantly higher detection rate of rarer C-bearing molecules such as C$_4$H$_2$. 
At a given accretion luminosity, molecular line luminosities are systematically lower in USco than in young disks, and the scaling relations with accretion luminosity differ between the two populations.
Moreover, we find that about half of the older disks, preferentially the millimeter faint, and likely more compact disks, have observable mass ratios of C- to O-bearing molecules that are higher than the maximum values in the young sample. These results point to reduced inner-disk molecular gas masses, cooler emitting layers, and higher inner gas C/O ratios in older disks, the latter being consistent with pebble drift. Taken together, our findings provide  evidence for chemical evolution of inner disk gas from young to older systems, with important implications for the accretion of primordial planetary atmospheres.
\end{abstract}


\section{Introduction}

Protoplanetary disks are the birthplaces of planets, and infrared (IR) spectroscopy is a useful tool to probe the gas  within a few au, i.e. inside the water snowline, where terrestrial planets assemble their cores and accrete their atmospheres \citep[e.g.,][]{Lee16,Johansen17,Henning13,Bean21}. Decades ago, the \textit{Spitzer} Space Telescope \citep{Werner04} obtained medium-resolution IR spectra for more than a hundred disks spanning a wide range of stellar masses, from more massive than the Sun down to brown dwarfs \citep[e.g.,][]{Pontoppidan10,Salyk11,CarrNajita11,Pascucci2013}. Today, the James Webb Space Telescope (JWST) \citep{Gardner23} delivers higher-resolution, more sensitive spectra in a similar spectral region, enabling deeper insight into the physical and chemical conditions of planet-forming regions \citep[e.g.,][]{Kamp23,Henning24,Arulanantham25}.

Mid-IR spectroscopy so far has mainly targeted young systems ($\lesssim 3$ Myr) and identified three major trends: 
1. High accretion luminosities ($L_{\rm acc}$) enhance all molecular emission \citep[e.g.,][]{Salyk11,Banzatti20}, pointing to a major role for UV-driven gas heating associated with accretion \citep[e.g.,][]{Dullemond07,Gorti08,Gorti09,Meijerink12,Walsh15,Najita17,Woitke18,Woitke24,Kanwar26}. 2. Disks around T Tauri stars (G, K, and early M types) typically show water-dominated spectra \citep[e.g.,][]{Pontoppidan14,Banzatti23,Temmink25}. In contrast, disks around very low-mass late M-type stars (VLMS, later than $\sim$M3),  are generally dominated by C$_2$H$_2$ and other hydrocarbons \citep[][]{Pascucci09,Pascucci2013,Tabone23,Arabhavi25,Long25,Grant25} at levels that cannot be explained by thermochemical models with solar C/O ratios \citep[e.g.,][]{Walsh15,Kanwar24a,Kanwar24}, while their stars photospheres have solar values \citep{Souto26}; and 
3. Millimeter-compact (generally faint) disks tend to have comparatively stronger water emission, with a subset showing enhanced cold ($T <400$\,K) water  \citep[e.g.,][]{Banzatti20,Banzatti23,Temmink25,Vlasblom25,Krijt25}. 

Importantly, the latter two trends suggest that the inner gas C/O ratio is not simply set by the snowlines of major volatiles, as proposed e.g. by \citet{Oberg11}. In that framework, inside the water snowline the C/O ratio is close to solar, while outside, as water vapor freezes onto grains, the gas becomes carbon-rich and the solids oxygen-rich.


To explain both the high C/O ratio of VLMS disks and enhanced cold water emission in some T~Tauri disks, 
a widely discussed framework is the pebble drift scenario \citep[e.g.,][]{Bosman17,Booth19}. In this scenario, the inner disk C/O ratio is set by the competition of  two processes: a) inward drift of icy pebbles, which sublimate and deliver oxygen at the water snowline \citep[($T \sim 170$\,K)][]{Ciesla06}, and b) advection of C-rich gas from the outer disk. At early times, efficient pebble drift produces a water-rich inner disk \citep[][]{Kalyaan21,Kalyaan23}; at later times, as the outer disk is depleted of pebbles and drift weakens, the inner disk transitions to a high-C/O state \citep[e.g.,][]{Booth19,Mah23,Sellek25}. As VLMS have closer in snowlines and, in the absence of traps, experience faster pebble drift, their disks transition to a higher inner C/O gas earlier than T~Tauri disks \citep[e.g.,][]{Pinilla13}. It has been also recently proposed that the release of organics from pebbles reaching  the so-call `soot' line ($T \sim 400$\,K) can also enhance the inner disk C/O ratio \citep{Kress10}.
The pebble drift scenario predicts that older and compact T~Tauri disks  may have run out of icy pebbles and entered a high C/O phase, while more extended disks can sustain a longer-lasting water supply and remain relatively low in C/O \citep[e.g.,][]{Kalyaan21,Mah24}. 



Recently, Carr \& Najita in prep. analyzed 12 disks in the $\sim 2-6$\,Myr IC~348 region with millimeter fluxes a factor of $\sim 5$ lower on average than younger disks. Based on the MIR fluxes of C- and O-bearing species,  they find no difference in the C/O ratios of the slightly older and young samples.
Moving to even older disks, in this work we present JWST/MIRI spectra of 14 disks in the Upper Scorpius (USco) association $\gtrsim 5$\,Myr. We combine our sample with  10 USco disks from the AGE-PRO sample (Raul et al. in prep.) and compare these older systems to a sample of young ($\sim 1-3$\,Myr) disks from the JDISCS Cycle~1 survey \citep{Arulanantham25} to search for evolution in the gas content.
The samples and observations are summarized in Section~\ref{sec:sample}, the analysis and fits to retrieve gas properties are described in Section~\ref{sec:results}. We discuss our results in Section~\ref{sec:dis} and provide a summary in Section~\ref{sec:summary}.

\section{Samples, Observations, and Data Reduction\label{sec:sample}} 
Here we describe the sample selection and observational strategy for the main USco sample (U2970), along with the sample selection for an additional set of USco sources (U3034, Raul et al. in prep.) and for a younger comparison sample \citep[JDISCS C1][]{Arulanantham25}, both of which are included in the discussion.
\subsection{Samples and Observations}
\begin{figure*}[htb!]
    \centering
	\includegraphics[width=0.90\textwidth]{ 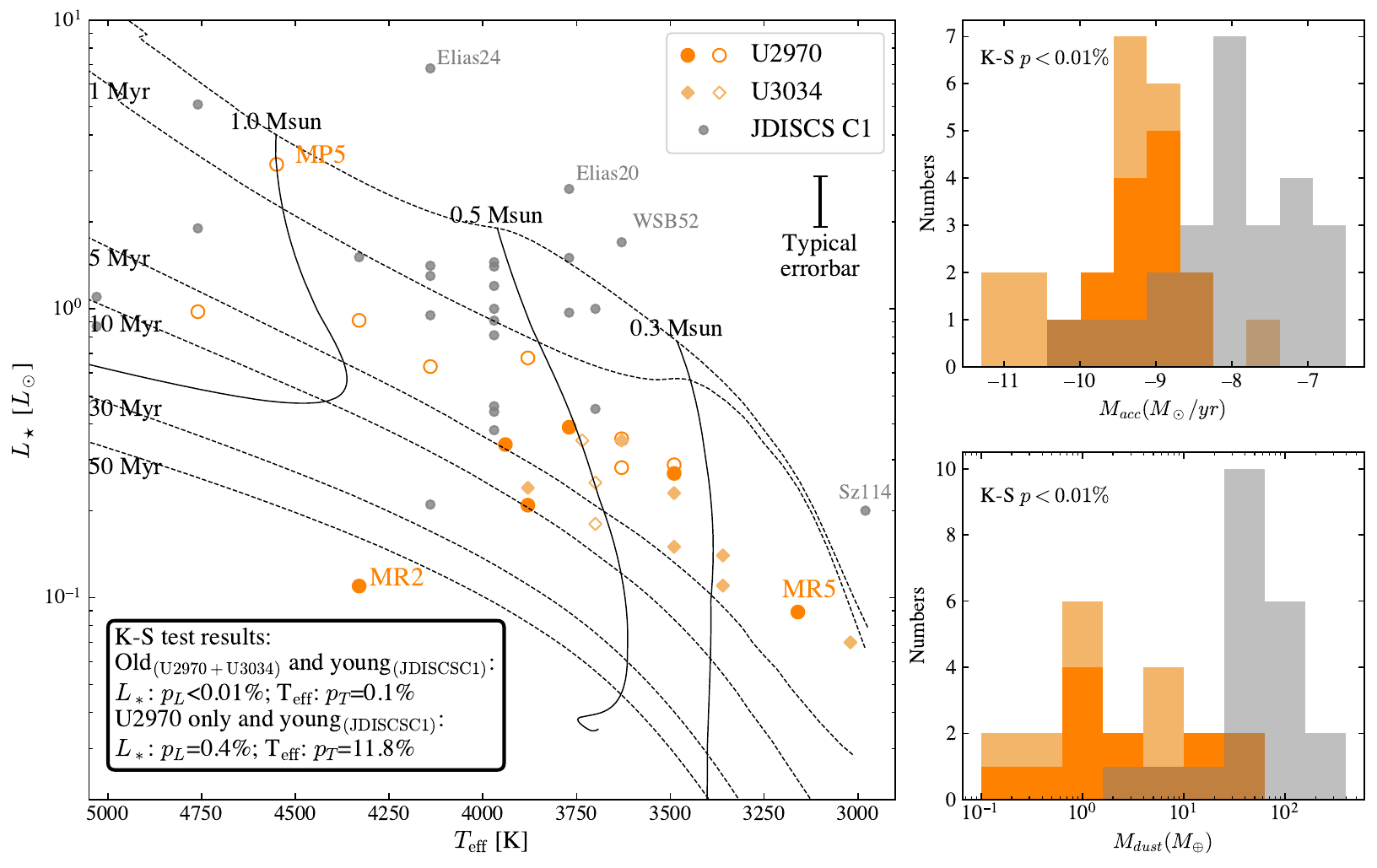}
    \caption{Left panel: HR diagram for the USco samples (U2970 in orange and U3034 in light brown), compared with younger sources from the JDISCS Cycle 1 survey \citep[gray,][]{Arulanantham25}. MR and MP disks (see Section~\ref{sec:res:overview} for the classification) are denoted with full and empty symbols, respectively. The results of the K-S tests for $L_*$ and $T_{\rm eff}$ between the young JDISCS sample and older USco samples are also shown. 
    The whole USco sample is significantly lower in $L_*$ and $T_{\rm eff}$ (with K-S $p<5\%$), but only U2970 (both MP and MR sub-samples) has similar $T_{\rm eff}$ as the younger sample (with K-S $p>10\%$). MR2 is highly inclined and its $L_{*}$ is underestimated \citep{Fang23}. 
    MP5 and MR5 are binary candidates in our JWST/MIRI cubes (see App.~\ref{app:binary} for details). 
    Right two panels: Accretion rate and dust mass distributions for each sample, with U3034 stacked on top of U2970. The older sample has significantly lower M$_{acc}$ and M$_{dust}$ than the young sample.
}
    \label{fig:overall:HR}
\end{figure*}
Our work uses data from the Cycle 2 General Observer (GO) program 2970 (PI: I. Pascucci, hereafter referred to as U2970, \citealt{jwst2024}) which were acquired with the MIRI \citep{Rieke15,Wright15,Wright23} MRS \citep{Wells15} instrument on JWST. U2970 comprises 14 T~Tauri stars with spectral types $\sim$M2-K2 ($0.3M_\odot < M_* < 1.5 M_\odot$) located in the $\sim$5-10 Myr older Upper Scorpius (USco) star-forming region \citep[e.g.,][]{Fang23,Ratzenbock23}, see Table~\ref{tab:para} for an overview. 
The 14 sources were selected from the 115 disks homogeneously analyzed in \citet{Fang23} to have: i) signatures of accretion (via e.g., the H$\alpha$ and/or [O {\scriptsize I}] 6300\,\AA lines, \citealt{Fang23}) 
and ii) ALMA millimeter (mm) continuum measurements \citep{Barenfeld16,Carpenter25}.
In order to explore how the inner disk chemical compositions might change with disk sizes, and because disk sizes correlate with millimeter fluxes \citep[e.g.,][]{Hendler20, Pinilla25}, we selected targets to cover three orders of magnitude in millimeter fluxes (F$_{\rm 0.89mm} \sim 1$ to $200$\,mJy).

U2970 was observed between 2024 July 21 and August 15. To enable high-contrast spectroscopy, each observation started with a target acquisition to place the source at the center of the field of view. The IFU data were taken in all three bands (SHORT, MEDIUM, and LONG) in all four channels to obtain complete spectral coverage ($\sim 4.9-28\,\mu$m). The exposure time for each target was set individually to reach a S/N of $\sim200$ at 16~$\mu$m, and a 4-Point dither pattern was used to provide robust sampling at all wavelengths and adequate point source separation in all channels. Along with the IFU exposures, simultaneous images were taken with F770W, F1000W and F1130W filters to improve the astrometric solution.

In our analysis, we also include the 10 USco disks from the JWST program GO~3034 (PI: K. Zhang, hereafter U3034, Raul et al. in prep.), which has a similar set-up as GO~2970, and reaches a S/N ratio of 100$-$200 at 22\,$\mu$m. These targets were selected from the ALMA Large Program AGE-PRO \citep{Zhang25}, with both mm-band continuum and $^{12}$CO (J=3-2) detections from prior shallower ALMA surveys \citep[][]{Barenfeld16}. We note that though the whole AGE-PRO sample was selected to cover spectral types from M3 to K6 \citep{Zhang25}, the 10 USco sources only cover M4.5 to M0 \citep[][and Raul et al. in prep.]{Agurto-Gangas25}, i.e. relatively later types than U2970.

To compare the old USco disks with a younger (1-3Myr) population, we selected sources from the JDISCS Cycle 1 sample, which consists of
17 disks from GO 1584 (PI: C. Salyk), 3 disks from GO 2025 (PI: K. Oberg), 3 disks from GO 1549 (PI: K. Pontoppidan) and 8 disks from GO 1640 (PI: A. Banzatti)\footnote{AS~209 is present in both GO 2025 and GO 1584.}. Hereafter, we call this sample JDISCS C1 \citep{Arulanantham25}. Disks selected in the JDISCS C1 sample have all been well studied at submillimeter wavelengths with ALMA \citep[see][for more details]{Arulanantham25}, and most of them (e.g., those from DSHARP-MIRI, GO 1584) have relatively high mm continuum  \citep[F$_{0.89}\gtrsim 100$\,mJy][]{Andrews18}. 
To allow for a more direct comparison, we only include the 2 K and M type stars located in young star-forming regions (Taurus, Lupus, and Ophiuchus), excluding four earlier type stars\footnote{HD~142666, HD~143006, HD~163296, MWC~480} and the known spectroscopic binary AS~205~S. 
The MIRI spectra of JDISCS C1 cover the same wavelength range, are observed with similar sensitivity and strategy as the U2970 and U3034 samples, and are extracted with similar data reduction pipelines \citep[e.g.,][and see Section~\ref{sec:datareduction} for more details]{Romero24,Pontoppidan24}. 

Fig.~\ref{fig:overall:HR} compares the three samples used in this work. The left panel shows their position in the Hertzsprung-Russell diagram, with bolometric luminosity ($L_*$) vs effective temperature ($T_{\rm eff}$) from \citet{Fang23,Agurto-Gangas25,Arulanantham25}, and evolutionary tracks from \citet{Feiden16}. The right panels show histograms of dust disk masses \citep[$M_{dust}$, ][]{Arulanantham25,Carpenter25,Agurto-Gangas25} and literature mass accretion rates from the Balmer jump or optical lines \citep[$\dot{M}_{\rm acc}$,][Empey et al. in prep.]{Manara20,Manara23,Fang23}. The rationale for using these $\dot{M}_{\rm acc}$ values, rather than the MIR H{\scriptsize I}$- \dot{M}_{\rm acc}$ conversion is provided in Appendix~\ref{app:Lacc}. 
To compare the distributions of stellar and disk properties between the young and older samples, we performed Kolmogorov–Smirnov (K-S) tests. The $p$ value gives the probability that the samples are drawn from the same parent population, with  $p<0.05$ corresponding to differences significant at the $2\sigma$ level. These values are reported in Fig.~\ref{fig:overall:HR}.
We find that, compared to the young sample, the older samples have lower stellar luminosities $L_*$ \citep[as expected from stellar evolution, e.g.,][]{Feiden16}, mass accretion rates $M_{\rm acc}$ \footnote{The trend persists when accounting for the known $M_{\rm acc} \propto M_*^2$ correlations \citep[e.g.,][]{Manara16}}  \citep[as noted in previous works on larger samples, e.g.,][]{Fang23,Delfini25}, and disk masses $M_{\rm dust}$ \citep[as already shown from initial ALMA surveys targeting entire nearby star-forming regions, e.g.,][]{Barenfeld16,Pascucci16}. The $T_{\rm eff}$ distribution of U2970 (both molecular rich and molecular poor subsamples, see Section~\ref{sec:res:overview} for the classification) is consistent with that of the younger sample, while U3034 shows lower $T_{\rm eff}$, indicating later spectral types.

\subsection{Data Reduction} \label{sec:datareduction}
All spectra from U2970 were extracted and wavelength-calibrated with the JDISCS pipeline described in \cite{Pontoppidan24}. 
The pipeline adopts the standard MIRI/MRS pipeline \citep{jwst2024} to stage 2b, and then calibrates the spectra with observed calibrator asteroid 515-Athelia spectra from GO program 3034 to remove fringes and maximize the S/N in channel 2, 3, 4. In channel~1, a standard star is used as a calibrator because the asteroid spectra have low S/N. Default extraction aperture radii are set as 1.4, 1.3, 1.2 and 1.1 times 1.22$\lambda/D$ for MRS channel 1 through 4, respectively. In all cases, the extraction apertures are kept the same between the source and the calibrator to remove all PSF complexities and improve the spectro-photometric precision. The latest version of the pipeline which is used for our sample now includes modeling of the asteroid with a wavelength-dependent emissivity benchmarked to photometric standards. 

Upon inspection of the MIRI channel 1 cubes, we discovered
four binary candidates with close separations (see Table~\ref{tab:para}). 
To assess the influence and potential contamination from the companions, we first extracted spectra using a larger aperture\footnote{We use $3\times1.22\lambda/D$ for channel 1 and $2\times1.22\lambda/D$ for channel 2, ensuring that more than 99\% of the flux from both components is included.}, so that emission from both stars is captured. We then performed simultaneous PSF fitting of the two components to determine the flux ratio between the primary and secondary at each wavelength. In three out of the four systems, one component shows a photospheric-like spectrum and therefore does not affect the spectral line features. The exception is J16141107, where contaminated features are present; 
this case is excluded from our current analysis and deferred to future work (see Appendix~\ref{app:binary}). 
Because the PSF fitting introduces significant noise and because the companion contributes only photospheric emission that does not affect the disk line features in our wavelength range of interest (11-19\,$\mu$m, where the combined spectrum is disk-dominated), we present and analyze only the combined spectra in the main text. 
The separated spectra, along with the associated uncertainties, are discussed in detail in Appendix~\ref{app:binary}.

The data reduction for U3034 sample is done as for U2970 (Raul et al. in prep.), while for JDISCS C1 the latest modeling of the asteroid has not been implemented, which will result in a relative flux difference of up to $\sim10\%$ in the wavelengths encompassing the major molecular lines ($\sim 10 -19\,\mu$m) (for more details, see Raul et al in prep. for U3034 and \citet{Arulanantham25} for JDISCS C1). 


\begin{deluxetable*}{cc|ccccccccc@{\hskip 1.5pt}c@{\hskip 1.5pt}c@{\hskip 1.5pt}c@{\hskip 1.5pt}c@{\hskip 1.5pt}c@{\hskip 1.5pt}c}
\tablecaption{Stellar and Disk Properties \label{tab:para}}
\tablewidth{0.99\textwidth}
\tablehead{
\multirow{2}{*}{ID}& \multirow{2}{*}{2MASS ID} & SpT & d &$M_*$  &  log$_{10}$($L_*$)  & log$_{10}$($\dot{M})$ & F$_{0.89}$ & n$_{13-26}$ & R$_{90}^{c}$& \multirow{2}{*}{\rotatebox[origin=c]{90}{H$_2$O$^*$}} & \multirow{2}{*}{\rotatebox[origin=c]{90}{C$_2$H$_2$}}  & \multirow{2}{*}{\rotatebox[origin=c]{90}{HCN}} &\multirow{2}{*}{\rotatebox[origin=c]{90}{CO$_2$}}&\multirow{2}{*}{\rotatebox[origin=c]{90}{C$_4$H$_2$}}  & \multirow{2}{*}{\rotatebox[origin=c]{90}{Ne$^*$}} & \multirow{2}{*}{\rotatebox[origin=c]{90}{H$_2$$^*$}} \\
 & & & (pc)& ($M_\odot$)  & ($L_\odot$) & ($M_{\odot}$/yr)& (mJy) & & (au) & & & & & &  &
}
\startdata
\multicolumn{17}{c}{Molecular Poor (MP) disks (no H$_2$O, C$_2$H$_2$, HCN or CO$_2$ is detected) }\\
\hline 
MP1&J16042165-2130284& K2 & 144.6 & 1.25  & -0.01 & $<$-10.34 & 217.40& 1.64& 111 (54) & \myboxwhi{}& \myboxwhi{}& \myboxwhi{}& \myboxwhi{}& \myboxwhi{}& \myboxblk{}& \myboxblk{}  \\
MP2&J16064794-1841437& K9 & 155.8 & 0.56 & -0.17 & -9.42 & 119.05&  3.19& 86 (29) &  \myboxwhi{}& \myboxwhi{}& \myboxwhi{}& \myboxwhi{}& \myboxwhi{}& \myboxblk{}& \myboxblk{}  \\
MP3&J16052157-1821412& K4 & 148.9 & 1.05 & -0.04 & -8.96 & 45.43 & 1.92 & 47 (28) & \myboxwhi{}& \myboxwhi{}& \myboxwhi{}& \myboxwhi{}& \myboxwhi{}& \myboxblk{}& \myboxblk{} \\
MP4&J16035767-2031055& K5 & 142.6 & 0.81 & -0.20 & -9.29 & 5.83 & 0.52 & $<$48 & \myboxwhi{}& \myboxwhi{}& \myboxwhi{}& \myboxwhi{}& \myboxwhi{}& \myboxblk{}& \myboxblk{}  \\
MP5$^a$&J16141107-2305362& K9 & 138.8 & 0.67  & -0.68 & -9.04 & 5.05  & -1.29 & $<$48 & \myboxgray{}{}& \myboxwhi{}& \myboxwhi{}& \myboxgray{}{}& \myboxwhi{}& \myboxblk{}& \myboxblk{}  \\
MP6$^a$&J16062196-1928445& M1 & 142.0 & 0.42  & -0.45 & -8.90 & 4.87  & 1.25 & $<$47 &\myboxwhi{}& \myboxwhi{}& \myboxwhi{}& \myboxwhi{}& \myboxwhi{}& \myboxblk{}& \myboxblk{}  \\
MP7&J16111534-1757214& M1 & 135.3 & 0.40 & -0.55 & $<$-9.69 & $<$0.65  & -1.18& $<$27 & \myboxwhi{}& \myboxwhi{}& \myboxwhi{}& \myboxwhi{}& \myboxwhi{}& \myboxwhi{}{}& \myboxgray{} \\
\hline 
\multicolumn{17}{c}{Molecular Rich (MR) disks}\\
\hline 
MR1&J16142029-1906481& K9 & 138.8 & 0.67  & -0.68 & -9.04 & 41.45 & -0.19& 19 (53) & \myboxblk{}& \myboxwhi{}& \myboxwhi{}& \myboxwhi{}& \myboxwhi{}& \myboxblk{}& \myboxblk{}  \\
MR2$^b$&J16075796-2040087& K4 & 135.9 & 0.71 & -0.96 & -8.96 & 23.9  & -0.44 &9 (25) & \myboxblk{}& \myboxblk{}& \myboxblk{}&  \myboxwhi{}& \myboxwhi{}& \myboxblk{}& \myboxblk{} \\
MR3&J16153456-2242421& M0 & 136.9 & 0.48 & -0.41 & -8.68 &  12.84  & 0.85 &13 (25) & \myboxblk{}& \myboxwhi{}& \myboxblk{}& \myboxwhi{}& \myboxwhi{}& \myboxblk{}& \myboxblk{} \\
MR4&J16123916-1859284& M2 & 134.7 & 0.33  & -0.54 & -8.69 & 9.45 & -0.50 & 27 (28) & \myboxblk{}& \myboxblk{}& \myboxblk{}& \myboxblk{}& \myboxblk{}& \myboxblk{} & \myboxblk{} \\
\multirow{2}{*}{MR5$^a$}&\multirow{2}{*}{J16120505-2043404}& M4$^{a}$ & \multirow{2}{*}{122.5} & - & -1.1 & -9.50 & 3.47  & -0.97& $<$24 & \myboxblk{}& \myboxblk{}& \myboxblk{}& \myboxblk{}& \myboxblk{}&\myboxwhi{} & \myboxblk{}  \\
& & M1.5 & & - & -0.69 &  - & -  & - & - \\
MR6$^a$&J16153220-2010236& M2 & 142.0 & 0.31 & -0.57 & -9.13 & 1.92 & -0.08 &$<$30 & \myboxblk{}& \myboxblk{}& \myboxblk{}& \myboxblk{}& \myboxblk{}& \myboxblk{}& \myboxblk{} \\
MR7&J16064385-1908056& K8 & 145.3 & 0.69 & -0.47 & -9.95 & $<$0.75 & -0.90 &$<$29 & \myboxblk{}& \myboxblk{}& \myboxblk{}& \myboxblk{}& \myboxblk{}& \myboxblk{}& \myboxblk{} \\
\hline
\multicolumn{2}{c}{References:} &1 &1 &1 &1 &1 &2 &3 & 2, 4 \\
\enddata
\tablerefs{1.\citet{Fang23}; 2. \citet{Carpenter25}; 3. This work; 4. \citet{Pinilla25}}
\tablecomments{$^*$For each molecule, a black square indicates detection, a gray square indicates a tentative detection, while a white indicates a non-detection. The detection of H$_2$O is determined by the hot and warm single lines at $\sim$17.32 and $\sim$17.51\,$\mu m$ \citep[see Table~1 in][]{Banzatti25}, Ne is determined by the [Ne {\scriptsize II}] forbidden line, and H$_2$ is determined by the H$_2$ S(1) transition.\\
a. The four sources are identified as binary candidates according to our JWST/MIRI channel 1 datacubes. According to coordinates cross-match between our JWST data and the ALMA images, the disk emission of MR5 and MR6 in our JWST spectra correspond to the mm emission from the `additional source' in \cite{Carpenter25} and have a candidate companion that does not contaminate their infrared spectra (note that for MR5 it is the secondary that has a disk, see Appendix.~\ref{app:binary}).
However, the spectrum of MP5 is contaminated by the secondary component, and the tentative detections of H$_2$O and CO$_2$ are likely false detections due to the contamination. Details about the binary candidates are discussed in Appendix~\ref{app:binary}. \\
b. MR2 is a highly inclined disk with a high accretion rate and a jet but obscured photosphere. Its spectral type and luminosity are more uncertain \citep{Fang23}. \\
c. The R$_{\rm 90}$ values are from visibility fittings \citep{Pinilla25} which obtain  a resolution up to 3 times better than the beam sizes \citep{Sierra25}. The values in brackets (or with `$<$' for unresolved disks) are the half beam sizes of each observation from \cite{Carpenter25}. All the MR disks are not or only barely resolved (R$_{90}$ smaller than beam sizes), thus here we use the F$_{0.89}$ as an indicator of the disk size \citep[e.g.,][]{Hendler20,Pinilla25}.
}
\end{deluxetable*}

\section{Analysis and Immediate Results for the U2970 sample \label{sec:results}
}

\subsection{Overview of the spectra and continuum subtraction \label{sec:res:overview}}
\begin{figure*}[htb!]
    \centering
	\includegraphics[width=0.90\textwidth]{ 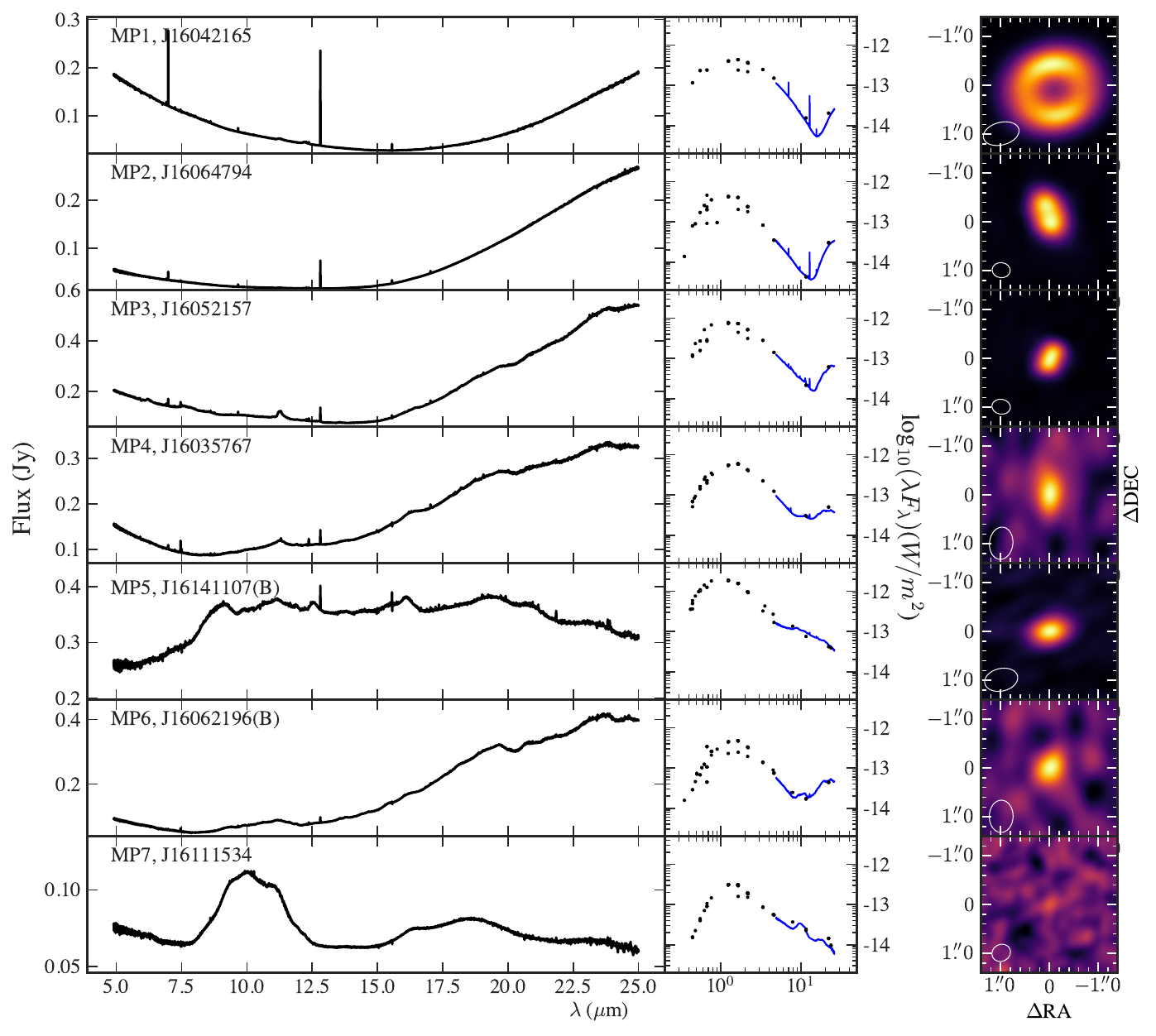}
    \caption{MIRI spectra and SEDs for the molecular poor (MP) disks in U2970, ordered by millimeter flux from high to low. `(B)' denotes binaries. The ALMA images are from \citet{Carpenter25}, centered at the corresponding source. Most disks have a high IR index and lack the 10-$\mu$m silicate emission feature. 
    }
    \label{fig:overall:poor}
\end{figure*}

\begin{figure*}[htb!]
    \centering
	\includegraphics[width=0.90\textwidth]{ 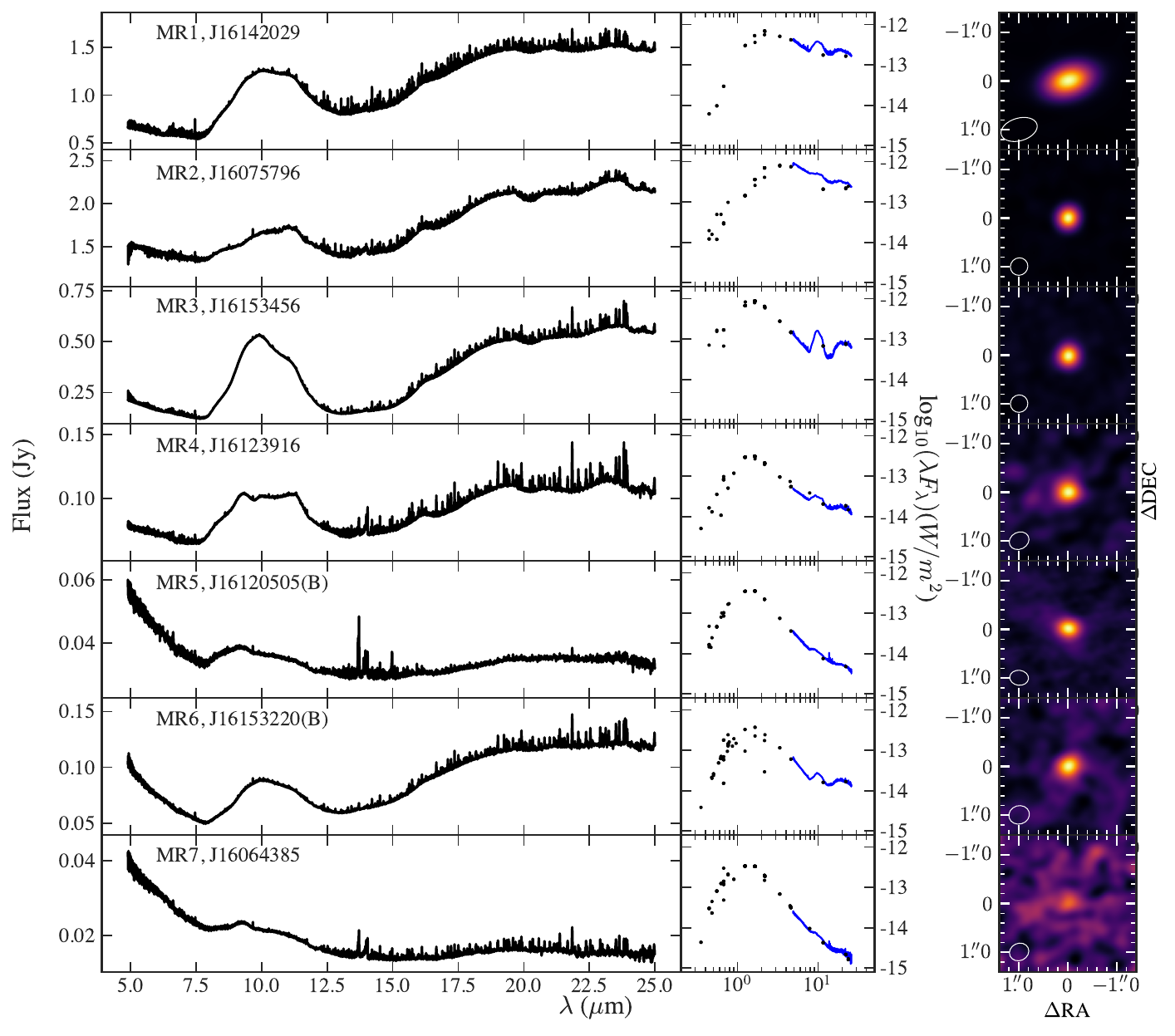}
    \caption{MIRI spectra and SEDs for the molecular rich (MR) disks in U2970, ordered by millimeter flux from high to low. `(B)' denotes binaries. The ALMA images are from \citet{Carpenter25}, centered at the corresponding sources. 
    }
    \label{fig:overall:rich}
\end{figure*}

The reduced U2970 spectra are shown in Figs.~\ref{fig:overall:poor} and \ref{fig:overall:rich}, together with the spectral energy distributions (SEDs) from \citet{Fang23} and the ALMA Band~7 millimeter continuum images from \citet{Carpenter25}. To analyze the spectral lines from these disks, we subtract the continuum from all spectra following the procedure outlined in \citet{Pontoppidan24}. Briefly, we estimate the underlying continuum with an iterative median filter, adopting window sizes between 45 and 200 wavelength channels depending on the wavelength range and spectral coverage to best reproduce the continuum shape. For disks with relatively strong C$_2$H$_2$ and HCN emission (J16075796, J16123916, J16120505, and J16124385), we exclude the 13.3–15.0~$\mu$m region from the continuum fitting in order to avoid over-subtracting the broad features surrounding the main organic molecular emission features. During the fitting, we use the line-free regions identified by \citet{Banzatti25} to apply a small wavelength-dependent flux offset and thus better match the expected continuum level. We then smooth the resulting continuum with a third-order Savitzky–Golay filter.

Because water, C$_2$H$_2$, HCN and CO$_2$ are the four strongest and most commonly detected molecular emissions in disks \citep[e.g.,][]{Salyk11,Pascucci2013,Banzatti25,Arabhavi25,Grant25,Long25}, we classify disks lacking any of the main species in the continuum-subtracted MIRI spectra as molecular poor (MP) and exclude them from our chemical composition (C/O ratio) analysis. The remaining disks are classified as molecular rich (MR)\footnote{The definition is different from \citet{Mallaney26} that introduced it only for cavity disks. See Section~\ref{sec:dis:cav} for more details.}. Here, non-detections are defined such that the peak of each molecular feature (the hot and warm H$_2$O lines near $\sim 17.3~\mu$m and $\sim 17.5~\mu$m; \citealt{Banzatti25}, and the Q-branch for other molecules) is less than three times the spectral noise, where the noise is taken as the standard deviation of the surrounding line-free regions \citep[e.g.,][and Section~\ref{sec:res:mollinefluxes} for more details]{Grant25}.  In total, we classify 7 disks as MP and 7 as MR, with their spectra shown in Figs.~\ref{fig:overall:poor} and \ref{fig:overall:rich}, respectively. The detections and non-detections of the main molecular species are summarized in Table~\ref{tab:para}. Besides the main molecular species, we also indicate detections for one rarer C-bearing species (C$_4$H$_2$), the ionic [Ne\,{\scriptsize II}] line,
and the H$_2$ (S1) line, none of which is contaminated by water. 

Based on the classification and mm fluxes, we label the disks MP1-MP7 and MR1-MR7. Within each group, indices decrease with mm flux: MP1 and MR1 have the highest mm fluxes, while MP7 and MR7 have the lowest.
We note that 5 out of the 7 MP disks have a positive IR index n$_{13-26}$, implying reduced emission from sub-micron/micron dust grains in the inner disk \citep[e.g.,][]{Espaillat14}. Because most of the disks are not or only marginally resolved with  ALMA  (see Table~\ref{tab:para}), in this paper we use the mm fluxes, which are known to correlated with disk sizes, as a tracer of the amount and location of outer disk pebbles \citep[e.g.,][]{Tripathi17,Hendler20}.

The spectra from the other USco sample (U3034) are presented in Raul et al., in prep., while the JDISCS C1 spectra have been already published in \cite{Arulanantham25}. Following our classification, the U3034 sample includes 3 MP (classified as `Molecular-Absent' in Raul et al. in prep.) and 7 MR disks, 
while in JDISCS C1 all the 25 young systems have water detections \citep{Arulanantham25}, hence are classified here as MR.





\subsection{LTE slab model fitting \label{sec:LTEfitting}}
For the MR disks, we analyze the inner disk molecular emission with slab models assuming the level populations of all species are in local thermodynamic equilibrium (LTE). The line broadening is assumed to be thermal (i.e., turbulence broadening is set to zero). In this way, models depend on three free parameters: column density $N$, gas temperature $T$, and emitting area $A$. These simple models have been used to reproduce dozens of mid-IR spectra \citep[e.g.,][]{CarrNajita11,Pascucci2013, Liu2019,Kamp23,Romero24,Gasman25,Banzatti25}.

We adopt the python package \texttt{iris} with spectroscopic data from the HITRAN database \citep{HITRAN22} to fit our data. \texttt{iris} includes a treatment of optical depth for line overlap \citep[see e.g.,][for details]{Tabone23,Romero24} and uses the Bayesian nested sampling Python package \texttt{dynesty} where random-walk sampling is adopted to explore the parameter space. The stopping criterion is set by the change in the remaining evidence ({\it Z}, marginal likelihood) when $\Delta \log Z \leq 0.01$. The spectra generated by iris are convolved to the JWST resolution which is wavelength-dependent. Our modeling focuses on the 11-19 $\mu$m range, encompassing emissions from all key molecules for inner disk chemistry analysis: hot and warm water, CO$_2$, and C-bearing species such as C$_2$H$_2$ and HCN (see Table~\ref{tab:para}).
Water emission spans nearly the entire 11-19\,$\mu$m range, overlapping with many other molecular and ionic emission features. Therefore, we first fit water emissions and then fit the C-bearing molecular lines after subtracting the water models. The detailed fitting process happens in three main steps as detailed below. 

\begin{figure*}[htb!]
    \centering
	\includegraphics[width=0.99\textwidth]{ 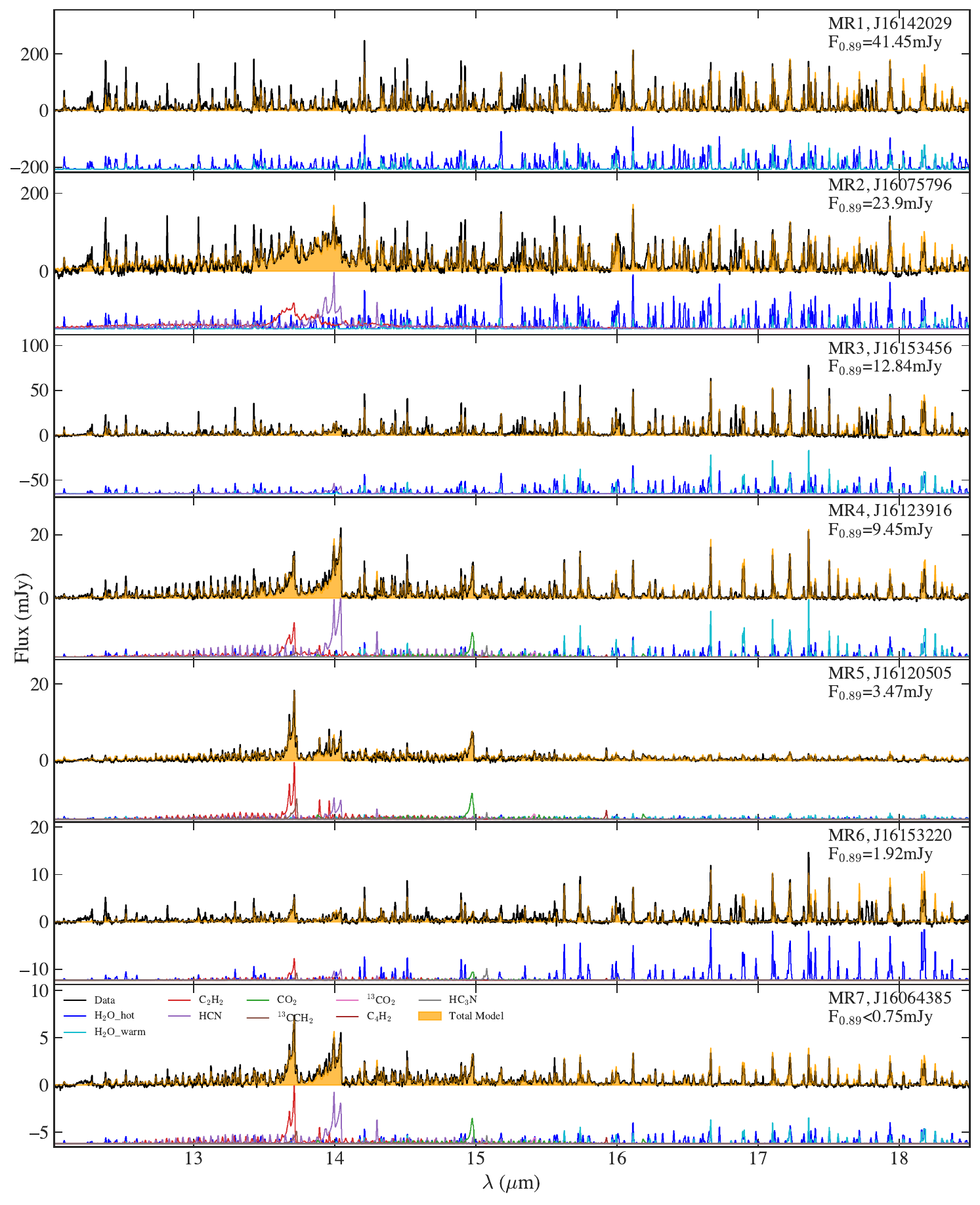}
    \caption{Continuum subtracted spectra and models for the MR disks in U2970, ordered from top to bottom by decreasing mm flux. The lower mm-flux disks show more C-rich spectra.
    }
    \label{fig:model:rich}
\end{figure*}

{\bf Step 1: Fitting of water lines.} 
We start by fitting H$_2$O with two temperature components \citep[hot and warm, e.g.,][]{Romero24}. For all disks we focus on the wavelength ranges between 11.5-12.2\,$\mu$m and 15.6-18.6\,$\mu$m to avoid the 13-15\,$\mu$m region that has many overlapping C-bearing molecular lines. 
We adopt uniform priors for emitting area and column density, with $\log_{10} A = \mathcal{U}(-3, 3)$\,au$^2$ and $\log_{10} N = \mathcal{U}(16, 20)$\,cm$^{-2}$. The prior of gas temperature $T$ is set as a normal distribution centered around 800\,K for the hot component and 400\,K for the warm component \citep[e.g.,][]{Romero24}, both with a standard deviation of 200\,K. We use the median value of the fitted posterior distribution as the best-fit value for each parameter. For source MR6, the two-component water fit failed to converge. Tests with one and three components showed that a single warm ($300\,\mathrm{K}< T < 600\,\mathrm{K}$) water component provides a good fit (see Table~\ref{tab:fitting} and Fig~\ref{fig:model:rich}). Following the U3034 analysis (Raul et al. in prep.), we classify the $T > 600\,\mathrm{K}$ component of water as hot, and $300\,\mathrm{K}< T < 600\,\mathrm{K}$ component as warm. 

{\bf Step 2: Fitting of the main C-bearing species}. For all disks, we first subtract from the spectrum our best-fit water models.
Because emission from C$_2$H$_2$, HCN, and CO$_2$ and their commonly detected isotopologues ($^{13}$CCH$_2$ and $^{13}$CO$_2$) are close in wavelength with each other (see the lower panels of Fig.~\ref{fig:model:rich}), we fit all of them together to account for overlapping emission. To simplify the fitting, we assume the isotopologues have the same emitting regions as their main species (i.e., the temperature T and emitting area A are the same), while allowing the column density to change \footnote{We do not fix the isotope ratio as emission from the main isotopologue could be optically thick (see App.~\ref{app:sec:tau}).}. 
We restricted the fitting range to 12.9-16.25\,$\mu$m to cover the main P, Q and R branches for all these molecules. For each molecule, we adopt uniform priors for all the parameters, with $\log_{10} A = \mathcal{U}(-3, 3)$\,au$^2$, $\log_{10} N = \mathcal{U}(12, 22)$\,cm$^{-2}$ and $\log_{10} T = \mathcal{U}(2.0, 3.2)$\,K. These priors are set to be very broad to ensure that the parameter space is fully explored. 
When the model does not find a solution for a specific molecule, we exclude that molecule, and refit the spectrum with the remaining ones. We then subtract the best fit model and evaluate if the excluded molecule is detected \citep[peak of the line $\ge 3\,\sigma$ of the adjacent continuum, e.g., ][]{Grant25}. If detected, we refit that molecule alone with a new LTE model using the residual spectrum obtained after subtracting the multi-species best-fit model. 
Otherwise, we report an upper limit on the column density $N$ (see Table~\ref{tab:fitting}). We derive these limits by fixing $T$ and $A$ to the best-fit values of the warm H$_2$O component for the same disk, and then determining the maximum $N$ that remains consistent with the spectral noise.
\footnote{We note that our fixed temperature is set to be the same as another molecule in the same disk to give a better estimate of the emitting mass ratio (see Section~\ref{sec:dis:COratio:emitmass}). On the other hand, Raul et al., in prep. fixed the temperature to be the average of the same molecules detected in other disks in their sample. This leads to discrepancy on the estimated emitting masses, and in our analysis, we recalculated the upper limits for the whole sample to estimate the C/O ratios consistently. 
} 
Among all MR disks, MR1 shows only water emission, MR2 lacks CO$_2$, and MR3 has no detections of C$_2$H$_2$ and CO$_2$.  The results of the fitting are listed in Table~\ref{tab:fitting} with non-detections indicated as $-$. The best fit models compared with the data are shown in Fig.~\ref{fig:model:rich}.


{\bf Step 3: Fitting of rarer C-bearing species.} Besides the main species, we also checked for other rarer, less abundant C-bearing molecules including CH$_4$, C$_2$H$_4$, C$_2$H$_6$, C$_6$H$_6$, HC$_3$N, C$_3$H$_4$, C$_4$H$_2$, CH$_3$. We report detections of HC$_3$N and C$_4$H$_2$ in all the relatively millimeter faint disks (F$_{0.89}<$10\,mJy, MR4 to MR7). To fit these molecules, we first subtract the best fit model of the main species, and adopt the same uniform prior for all parameters as discussed above. Because the emission of HC$_3$N and C$_4$H$_2$ are not blended with each other, we fit them separately and restricted the fitting range as 15.0-15.2\,$\mu$m for HC$_3$N and 15.8-16.0\,$\mu$m for C$_4$H$_2$, respectively. The fitting results are also shown in Table~\ref{tab:fitting} and Fig.~\ref{fig:model:rich}. 

We note that in most cases the column densities are degenerate with the emitting area (e.g., the HCN of MR2). To address this, we also use in Sect~\ref{sec:dis:COratio} the observable mass defined as the column density $N$ times the emitting area ($A$), see also \citet[][]{Arulanantham25}. We note that for MR2 the C$_2$H$_2$ and HCN emission in the 13.5-14.0\,$\mu$m region possibly produces a pseudo-continuum and is likely optically thick, see Fig.~\ref{fig:model:rich}. Since no additional information about the continuum is available in this wavelength range, the C$_2$H$_2$ and HCN best fit parameters for MR2 are less well constrained \citep[see Sz~114,][for a similar case]{Xie23}. We also note that for MR6, because the well-constrained C$_2$H$_2$ temperature (inferred from the suppressed secondary peak of the Q branch of C$_2$H$_2$ emission) is low, the best-fit model indicates a high C$_2$H$_2$ abundance. This is consistent with the absence of hot water, implying this disk is cool and relatively carbon rich, which is further supported by the detections of more complex C-bearing species (i.e., HC$_3$N and C$_4$H$_2$, see also Section~\ref{sec:dis:COratio}). 

\begin{deluxetable*}{ll|ccccccc}
\tablecaption{Best fit models of the U2970 sample. The fitted parameters shown here include column density (N), emitting temperature (T) and emitting area (A). 
\label{tab:fitting}}
\tablewidth{0.99\textwidth}
\tablehead{
\multicolumn{2}{c}{ID} & MR1 & MR2 & MR3 & MR4 & MR5 & MR6 & MR7 
}
\startdata
\multirow{3}{*}{\shortstack{H$_2$O hot \\($T > 600\,\mathrm{K}$)}} & $\log_{10}$N (cm$^{-2}$) & 17.97$\pm$0.04 & 17.49$\pm$0.04 & 17.49$\pm$0.03 & 17.77$^{+0.10}_{-0.09}$ & 19.78$^{+0.18}_{-0.25}$ & -- & 17.58$\pm$0.07\\
& T (K) & 870$\pm$20 & 980$\pm$30 & 850$\pm$30& 680$\pm$20 & 740$^{+60}_{-50}$& -- & 740$\pm$20 \\
& $\log_{10}$A (au$^2$) & -0.03$\pm$0.03 & -0.07$\pm$0.02 & -0.53$\pm$0.02 & -0.99$\pm$0.04 & -2.35$^{+0.11}_{-0.14}$ & -- & -1.46$\pm$0.03 \\
\hline
\multirow{4}{*}{\shortstack{H$_2$O warm\\($300\,\mathrm{K}< T < 600\,\mathrm{K}$)}}& $\log_{10}$N (cm$^{-2}$) & 17.55$^{+0.07}_{-0.08}$ & 18.47$^{+0.30}_{-0.26}$ & 15.99$^{+0.22}_{-0.67}$ & 16.49$^{+0.13}_{-0.15}$ & 18.18$^{+0.19}_{-0.17}$ & 16.92$^{+0.02}_{-0.03}$ & 16.73$^{+0.12}_{-0.16}$ \\
& T (K) & 490$\pm$10 & 350$\pm$20 & 490$\pm$10 & 440$\pm$10& 580$\pm$20& 575$\pm$10 & 480$\pm$10 \\
& $\log_{10}$A (au$^2$) & 0.8$\pm$0.02 & 0.73$\pm$0.07 & 1.48$^{+0.64}_{-0.2}$ & 0.78$^{+0.13}_{-0.10}$ & -1.53$\pm$0.03 & -0.08$\pm$0.02 & -0.29$^{+0.11}_{-0.07}$ \\
\hline
\multirow{3}{*}{C$_2$H$_2$}& $\log_{10}$N (cm$^{-2}$) & \color{gray}$<$14.3 & 18.3$\pm$0.1 & \color{gray}$<$12.5 & 16.59$^{+0.06}_{-0.05}$ & 16.78$^{+0.07}_{-0.06}$ & 17.06$^{+0.34}_{-0.29}$ & 16.06$^{+0.10}_{-0.11}$ \\
& T (K) & \color{gray}[490] & 1260$^{+190}_{-140}$ & \color{gray}[490] & 740$\pm$20 & 310$\pm$20 & 150$\pm$20 & 490$\pm$20 \\
&$\log_{10}$A (au$^2$)& \color{gray}[0.8] & -2.1$^{+0.08}_{-0.10}$ & \color{gray}[1.48] & -1.73$\pm$0.04 & -0.64$\pm$0.05 & 0.96$^{+0.42}_{-0.37}$ & -1.23$^{+0.09}_{-0.08}$ \\
$^{13}$CCH$_2$ & $\log_{10}$N (cm$^{-2}$) & -- & -- & -- & 16.17 & 16.29 & 15.12 & 15.52 \\
\hline
\multirow{3}{*}{HCN}& $\log_{10}$N (cm$^{-2}$) & \color{gray}$<$15.0 & 13.4$^{+1.6}_{-1.2}$ & 16.25$^{+0.12}_{-0.14}$ & 15.45$\pm$0.06 & 16.10$\pm$0.06 & 16.16$\pm$0.16 & 15.76$^{+0.13}_{-0.20}$ \\
& T (K) & \color{gray}[490] &1020$\pm$30  & 510$\pm$30 & 500$\pm$10 & 470$\pm$20 & 370$^{+50}_{-40}$&560$^{+30}_{-20}$ \\
& $\log_{10}$A (au$^2$) & \color{gray}[0.8] & 2.8$^{+1.2}_{-1.6}$ & -0.68$\pm$0.07 & 0.22$^{+0.06}_{-0.05}$ & -0.86$\pm$0.04 & -0.88$^{+0.13}_{-0.11}$ & -0.65$^{+0.40}_{-0.21}$ \\
\hline
\multirow{3}{*}{CO$_2$}& $\log_{10}$N (cm$^{-2}$) & \color{gray}$<$14.9 & \color{gray}$<$15.0 & \color{gray}$<$13.2 & 16.51$^{+0.10}_{-0.11}$ & 16.90$^{+0.10}_{-0.09}$ & 16.5$\pm$0.6 & 16.93$^{+0.38}_{-0.29}$ \\
& T (K) & \color{gray}[490] & \color{gray}[350] & \color{gray}[490] & 320$\pm$20 & 420$^{+20}_{-30}$ & 270$^{+100}_{-80}$ & 300$^{+50}_{-70}$ \\
& $\log_{10}$A (au$^2$) & \color{gray}[0.8] & \color{gray}[0.73] & \color{gray}[1.48] & -0.49$^{+0.07}_{-0.05}$ & -1.18$\pm$0.05 & -0.73$^{+0.55}_{-0.44}$ & -0.89$^{+0.16}_{-0.09}$ \\
$^{13}$CO$_2$& $\log_{10}$N (cm$^{-2}$) & -- & -- & -- & 15.26 & 15.63 & -- & 15.5 \\
\hline
\multirow{4}{*}{HC$_3$N}& $\log_{10}$N (cm$^{-2}$) & -- & -- & -- & 14.88$^{+0.20}_{-0.12}$ & 15.08$\pm$0.14 & 15.89$\pm$0.16 & 15.12$^{+0.34}_{-0.29}$ \\
& T (K) & -- & -- & -- & 250$\pm$40 & 170$^{+20}_{-10}$ & 160$\pm$10 & 160$\pm$20\\
& $\log_{10}$A (au$^2$) & -- & -- & -- & [0.0] &[0.0] &[0.0] &[0.0]   \\
\hline
\multirow{4}{*}{C$_4$H$_2$}& $\log_{10}$N (cm$^{-2}$) & -- & -- & -- & 14.83$^{+0.76}_{-0.72}$ & 15.12$^{+0.13}_{-0.14}$ & 14.8$^{+1.12}_{-0.73}$ & 14.9$^{+0.63}_{-0.57}$ \\
& T (K) & -- & -- & -- & 160$^{+70}_{-40}$ & 170$^{+20}_{-10}$ & 140$^{+70}_{-30}$ & 150$^{+60}_{-30}$\\
& $\log_{10}$A (au$^2$) & -- & -- & -- & [0.0] & [0.0] &[0.0] &[0.0]  \\
\enddata
\tablecomments{Parameters held fixed during the fit are indicated in brackets (`[ ]'). For disks without detections of the main species (C$_2$H$_2$, HCN, and CO$_2$), we report estimated upper limits (shown in gray) on the column density $N$, fixing $T$ and $A$ to the values of the best fit model of warm H$_2$O for the same disk. }
\end{deluxetable*}

\subsubsection{Molecular line fluxes \label{sec:res:mollinefluxes}}
With the fitted models discussed above, we calculate line fluxes for the U2970 sample as follows.
For the C-bearing molecules, because their main branches are blended with each other, we integrate the best-fit model over fixed wavelength ranges listed in Table~\ref{tab:fitting}. For C$_2$H$_2$, HCN, and CO$_2$, we adopt a common range of 12.0–16.0 $\mu$m, matching \citet[][]{Arulanantham25} and Raul et al., in prep. to compare with other samples. For HC$_3$N and C$_4$H$_2$, we use the same wavelength ranges as Raul et al. (in prep.), centered on the peak of their Q branches. 

For water, we choose the higher-, intermediate-, and lower-energy single lines identified in Table~1 of \citet{Banzatti25}. These are isolated lines with no contamination from other atomic or molecular emission lines, thus we directly integrate the continuum subtracted spectra within the same wavelength range as in previous studies and specified in Table~\ref{tab:fluxes}. 


To derive the total line fluxes and their associated uncertainties, we use the \texttt{specutils} and \texttt{astropy.nddata.StdDevUncertainty} packages with input of wavelengths ranges, fluxes and corresponding noise level. 
The noise level is estimated from the random noise (standard deviation) in line-free regions of the continuum-subtracted spectrum adjacent to each integrated feature. This standard deviation is then adopted as the uncertainty per wavelength over the range used for the line integration. 
Because the integration window for the main C-bearing molecules (C$_2$H$_2$, HCN, and CO$_2$) is much broader than the actual emission, the standard deviation is measured within a narrower region ($\pm 0.1~\mu$m) around the peak of each Q-branch to avoid overestimating the noise. The resulting fluxes and uncertainties are listed in Table~\ref{tab:fluxes}. 
In Table~\ref{tab:fluxes} and ~\ref{tab:Ionic_H2}, non-detections (fluxes below 1$\sigma$) are represented as 3\,$\sigma$ upper limits, denoted with a “$<$” symbol, while tentative detections (fluxes between 1\,$\sigma$ and 3\,$\sigma$) are shown in brackets.


\begin{deluxetable*}{c|ccc||cccc|cc}
\tablecaption{C- or O- bearing molecular line fluxes (Flux shown in 10$^{-15}$ erg / (s cm$^2$))}\label{tab:fluxes}
\tablewidth{0.99\textwidth}
\tablehead{
ID & C$_2$H$_2$& HCN& CO$_2$ & H$_2$O 6000K& H$_2$O 3600K &  H$_2$O 1500K (a) &  H$_2$O 1500K (b) &   HC$_3$N& C$_4$H$_2$ \\
\multicolumn{2}{l}{Range ($\mu$m)} & \multicolumn{2}{l}{[12.0,16.0]} & [17.317, 17.33] & [17.49,17.52] & [23.805, 23.83] & [23.88,23.91] & [15.05, 15.09] & [15.908,15.938] 
}
\startdata
MR1 & $<$4 &$<$4 & $<$4 & 6.56$\pm$0.17 & 12.0$\pm$0.3 & 9.0$\pm$0.4 & 10.0$\pm$0.5 & $<$1.0 & $<$0.8 \\
MR2 & 622$\pm$2 & 489$\pm$2 & $<$5& 6.0$\pm$0.3& 8.5$\pm$0.4 & 5.6$\pm$0.4 & 3.9$\pm$0.5 & $<$1.7 & $<$1.0  \\
MR3 & $<$0.5 & 23.9$\pm$0.2 & $<$0.5& 1.53$\pm$0.04 & 3.61$\pm$0.05 & 8.7$\pm$0.3 & 7.4$\pm$0.3 &$<$0.03 & $<$0.04  \\
MR4 & 32.70$\pm$0.02 & 38.23$\pm$0.02 & 8.17$\pm$0.02& 0.297$\pm$0.013& 0.970$\pm$0.019 & 2.24$\pm$0.09& 1.89$\pm$0.10 &0.76$\pm$0.03 & 0.10$\pm$0.03  \\
MR5 & 22.81$\pm$0.03 & 11.44$\pm$0.03 & 9.49$\pm$0.03 & 0.088$\pm$0.004& 0.128$\pm$0.006& 0.17$\pm$0.03& 0.16$\pm$0.03 & 0.294$\pm$0.010&0.335$\pm$0.008  \\
MR6 & 4.95$\pm$0.06 & 4.05$\pm$0.06 & 1.67$\pm$0.06& 0.273$\pm$0.011 & 0.742$\pm$0.016 & 1.30$\pm$0.05& 1.47$\pm$0.05 & 0.541$\pm$0.03 & 0.050$\pm$0.014 \\
MR7 & 10.26$\pm$0.02 & 11.38$\pm$0.02 & 2.94$\pm$0.02 & 0.096$\pm$0.004& 0.215$\pm$0.006& 0.175$\pm$0.013& 0.238$\pm$0.014& 0.134$\pm$0.007 & 0.086$\pm$0.005 
\enddata
\end{deluxetable*}





\begin{deluxetable*}{cc|ccccc|c}
\tablecaption{Ionic and H$_2$ line fluxes (Flux shown in 10$^{-15}$ erg / (s cm$^2$))}\label{tab:Ionic_H2}
\tablewidth{0.99\textwidth}
\tablehead{
ID& 2MASS ID & [Ar\,{\scriptsize II}] & [Ne\,{\scriptsize II}] & [Ne\,{\scriptsize III}] & H$_2$ (S3)& H$_2$ (S1) & \multirow{2}{*}{ T$_{\rm H_2}$ (K)}\\
\multicolumn{2}{c}{Range ($\mu$m)} & [6.98,6.99] &  [12.80,12.82] & [15.545,15.565] &[9.658, 9.670] &[17.025,17.045] 
}
\startdata                                                                                       
MP1& J16042165& 20.97$\pm$0.05& 15.99$\pm$0.02& 0.989$\pm$0.009 & 0.66$\pm$0.02  & 0.150$\pm$0.009 & 579\\        
MP2& J16064794& 2.44$\pm$0.03& 4.598$\pm$0.005& 0.530$\pm$0.005 & 0.460$\pm$0.007  &  0.319$\pm$0.009 & 404\\      
MP3& J16052157& 4.42$\pm$0.09& 5.88$\pm$0.02& 0.81$\pm$0.03 & 1.26$\pm$0.04  & 0.698$\pm$0.024 & 430\\    
MP4& J16035767& 1.16$\pm$0.05& 2.72$\pm$0.03& 0.48$\pm$0.03 & 0.20$\pm$0.03  &  0.473$\pm$0.022 & 301\\ 
MP5& J16141107& 1.63$\pm$0.18 & 3.82$\pm$0.06& 1.97$\pm$0.05 & (0.15$\pm$0.07)  & 0.41$\pm$0.05 &  295\\ 
MP6& J16062196& 0.52$\pm$0.05 & 1.56$\pm$0.03& 0.463$\pm$0.019 & 0.36$\pm$0.03  &   0.193$\pm$0.019 & 434\\    
MP7& J16111534& $<0.26$& $<0.07$& $<0.04$ & $<0.09$  & (0.038$\pm$0.013) & $<467$\\    
\hline                               
MR1& J16142029& $<7.6$& 13.8$\pm$0.7& 3.6$\pm$0.3 & 3.37$\pm$0.5  & (1.1$\pm$0.5) & 503\\  
MR2& J16075796& $<6.7$& 13.9$\pm$0.8& 5.4$\pm$0.4 & 9.37$\pm$0.5  & 3.28$\pm$0.6 & 496\\
MR3& J16153456& $<1.9$& 1.90$\pm$0.12& 0.91$\pm$0.05 & 1.19$\pm$0.13  & 0.73$\pm$0.11 &   418\\      
MR4& J16123916& $<0.64$& 0.19$\pm$0.06 & 0.26$\pm$0.04 & 0.21$\pm$0.03  & 0.141$\pm$0.015 & 410\\
MR5& J16120505& $<0.20$& $<0.07$& $<0.23$ & 0.090$\pm$0.019  & 0.126$\pm$0.008 & 340\\       
MR6& J16153220& $<0.40$& 0.41$\pm$0.03& 0.476$\pm$0.016 & 0.33$\pm$0.03  & 0.167$\pm$0.015 & 440\\  
MR7& J16064385& $<0.19$& 0.074$\pm$0.011& 0.126$\pm$0.015 & 0.079$\pm$0.011  & 0.059$\pm$0.007 & 396\\          
\enddata

\end{deluxetable*}

\subsection{H$_2$ and ionic lines} \label{sec:H2_ionic_fluxes}
In addition to the C- and O-bearing species mentioned above, we measure fluxes for the H$_2$ and ionic lines (i.e., [Ne\,{\scriptsize II}], [Ar\,{\scriptsize II}]), which are present in almost all the disks in our sample (see Table~\ref{tab:para}). 

The lines fluxes and errors are measured similarly to the molecular lines with similar classifications of detections and non-detections (see Section.~\ref{sec:LTEfitting} for details). 
For the MP disks, the line fluxes are estimated directly from integrating the continuum subtracted spectra within wavelength ranges encompassing the lines.
For the MR disks, we first subtract the best fit models of all detected molecular species (shown in Table~\ref{tab:fitting}) and apply the same method.

The results of the ionic and H$_2$ line fluxes are summarized in Table~\ref{tab:Ionic_H2}. MP7 has no ionic line detections and only a tentative detection of the H$_2$\,(S1) line. 
Along with the lack of other molecules and an upper limit on the accretion rate, MP7 is likely a more evolved low gas mass disk or maybe even a young debris disk.
For the MR disks (MR1-MR7) where water is always detected, the [Ar\,{\scriptsize II}] line is not detected in any of them but the upper limits are high. A possible reason for this is that the [Ar\,{\scriptsize II}] line is blended with high-excitation water lines, which are likely out of LTE at these short wavelengths \citep[e.g.,][]{Banzatti25} and thus are not well accounted for by our LTE slab models. [Ne\,{\scriptsize II}], on the other hand, is blended with the R-branch of C$_2$H$_2$. Though the R branches are weak and well produced by our LTE slab models, we note that the flux values of [Ne\,{\scriptsize II}] of MR4-MR7 (with strong C$_2$H$_2$ emissions) can also be uncertain.
In Appendix~\ref{app:sec:dis:ionization}, we use ionic line ratios to show that, for most of the U2970 disks, the disk surface is ionized by X-ray or soft EUV photons.

The H$_2$ S(1) and S(3) lines are detected in all of the U2970 disks except MP7. 
These two H$_2$ populations are likely to be in LTE because their critical densities are low \citep{Mandy93} compared with the typical density of protoplanetary disks in the H$_2$ emitting region 
\citep[e.g.,][]{Woitke2009A&A...501..383W}. With the further assumption that the emission is optically thin, we use eq.~1 in \cite{Lahuis07} and derive the H$_2$ temperature based on these strong and non-contaminated H$_2$  lines, see Table~\ref{tab:Ionic_H2}. We note that most of the temperatures are relatively low ($\lesssim 500$\,K), consistent with temperatures near the disk surface \citep[e.g.,][]{Kamp04,Woitke18}. 


\section{Discussion: Comparing young and older disks \label{sec:dis}}

In the following sections we combine the older USco samples (U2970 $+$ U3034, hereafter combined USco sample) and compare them with the young disk sample (JDISCS C1) to explore evolutionary trends in the inner disk chemistry. 

In Section~\ref{sec:dis:linefluxes}, we show molecular line detection rates and how the line luminosities depend on stellar accretion. We note that all three samples are selected based on previous ALMA instead of mid-IR studies, and thus the detection rates are not biased by sample selections at infrared wavelengths. We also discuss how cool-to-hot water ratios may relate to the amount of pebbles in the outer disk (proportional to the millimeter flux) and the pebble drift.
In Section~\ref{sec:dis:cav}, we discuss how disk cavities influence the molecular emission, and in Section~\ref{sec:dis:COratio} we explore if there is any evolution in the C/O ratios of the molecular-rich disks.

\subsection{Molecular line detections and luminosities} \label{sec:dis:linefluxes}

\subsubsection{Detection rates} \label{sec:dis:linefluxes:detectionrates}
\begin{figure}[htb!]
    \centering
	\includegraphics[width=0.49\textwidth]{ 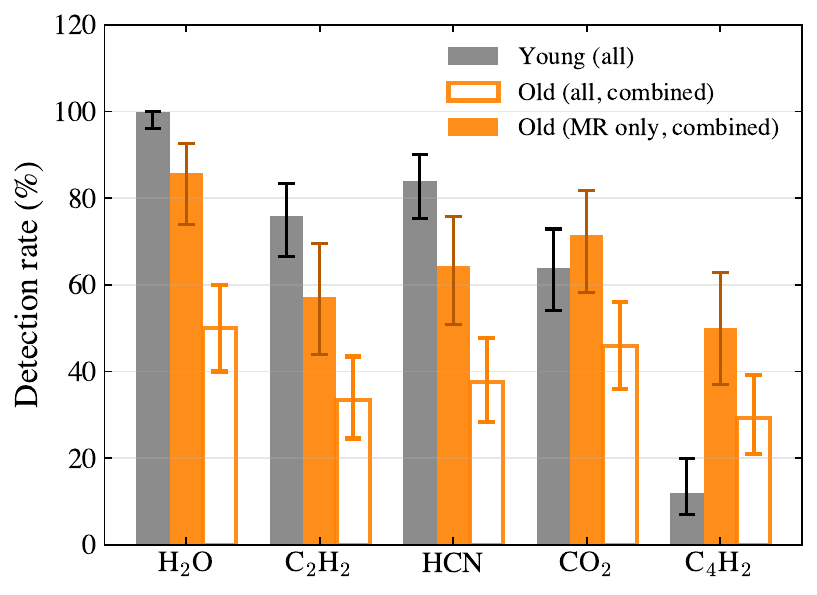}
    \caption{Detection rates for each molecule and sample. 
    For the full USco sample, detection rates of all major molecules are significantly lower relative to young disks, except for CO$_2$. On the contrary, the detection rate of the rarer C-bearing molecule C$_4$H$_2$ is higher in USco, hinting at elevated gas-phase C/O ratios. 
    }
    \label{fig:molecule:detectionrate}
\end{figure}

We compare the molecular detection rates for the young and older samples in Table~\ref{tab:detection_rates} and Fig~\ref{fig:molecule:detectionrate}. 
Because the HC$_3$N transition overlaps with H$_2$O emission, we instead opt for C$_4$H$_2$ as a representative rarer C-bearing species  \citep[e.g.,][the Q band of C$_4$H$_2$ is affected the least by water emission]{Grant25,Banzatti25}.
However, C$_4$H$_2$ detection rates for the JDISCS C1 sample are not reported in \citet{Arulanantham25}. Here, we estimate the line fluxes and uncertainties from the published continuum-subtracted spectra, applying the same method described in Section~\ref{sec:res:mollinefluxes} and same integration region as in Table~\ref{tab:fluxes}\footnote{The noise is estimated in the adjacent [15.902, 15.908]$\mu$m line-free region}. 
Only 3 out of 25 disks (GO~Tau, DoAr~33, and Elias~20) have C$_4$H$_2$ detections. We note, however, that the upper limits for JDISCS C1 are typically higher because the noise is larger.
As roughly half of the older sample shows little or no molecular emission, we also report detection rates for the MR subset only. The uncertainties on the detection rates are  the Wilson score intervals as implemented in \texttt{statsmodels.stats.proportion.proportion\_confint}. 
The key findings are:

(a) All young disks are MR, consistent with ubiquitous H$_2$O detections in disks around K and early M-type stars. In contrast, roughly half of the old disks are MP despite having similar stellar properties to the young sample. This suggests a decrease in the inner-disk molecular gas content at the old age of USco.

(b) The full USco sample shows lower detection rates than the young sample for H$_2$O, HCN, and C$_2$H$_2$, but maintains a similar detection rate  for CO$_2$. Because the 15\,$\mu$m CO$_2$ band traces cooler gas that may arise from deeper layers and/or larger radii in the disk \citep[e.g.,][]{Woitke18,Temmink24,Vlasblom24,Vlasblom25}, this sustained CO$_2$ detection rate is consistent with cooler inner disks at older ages. This cooling is likely linked to lower accretion luminosity ($L_{\rm acc}$), see also Sect.~\ref{sec:dis:line_cont}. 



(c) In contrast to the general decline in molecular detections, the C$_4$H$_2$ detection rate is higher in both the full or MR  USco sample.  This may result from elevated gas-phase C/O ratio (see also Sect.~\ref{sec:dis:COratio}).  
We caution, however, that the young sample has a brighter continuum and larger noise, leading to less stringent upper limits.

Taken together, these findings suggest a decline in the inner-disk molecular gas mass and/or decrease in gas temperature with age, and hint at a potential increase in the inner gas-phase C/O ratio.

\begin{deluxetable}{c|ccc|c}
\tablecaption{Detection rates of various molecules from different samples}\label{tab:detection_rates}
\tablewidth{0.49\textwidth}
\tablehead{
& \multicolumn{3}{c}{Old} & Young\\
&U2970 & U3034 & Combined & JDISCS C1
}
\startdata
H$_2$O& 7/14 (50\%) &5/10 (50\%)& 12/24 (50\%)& 25/25 (100\%)\\
C$_2$H$_2$& 5/14 (36\%) & 3/10 (30\%)&8/24 (33\%)& 19/25 (76\%)\\
HCN& 6/14 (43\%) &3/10 (30\%)&9/24 (38\%)& 21/25 (84\%)\\
CO$_2$& 4/14 (29\%) &7/10 (70\%)&11/24 (46\%)& 16/25 (64\%)\\
C$_4$H$_2$& 4/14 (29\%) &3/10 (30\%)&7/24 (29\%)& 3/25 (12\%)\\
\hline
\multicolumn{5}{c}{MR only} \\
\hline
H$_2$O& 7/7 (100\%) & 5/7 (71\%)& 12/14 (86\%)& 25/25 (100\%)\\
C$_2$H$_2$& 5/7 (71\%) & 3/7 (43\%)&8/14 (57\%)& 19/25 (76\%)\\
HCN& 6/7 (86\%) &3/7 (43\%)&9/14 (64\%)& 21/25 (84\%)\\
CO$_2$& 4/7 (57\%) &6/7 (86\%)&10/14 (71\%)& 16/25 (64\%)\\
C$_4$H$_2$& 4/7 (57\%) &3/7 (43\%)&7/14 (50\%)& 3/25 (12\%)\\
\enddata
\tablecomments{Detection rates for U2970 and C$_4$H$_2$ of JDISCS C1 are from this work, for U3034 from Raul et al. in prep., and for the main species of JDISCS C1 are from \citet{Arulanantham25}. Tentative detections are included.}
\end{deluxetable}

\subsubsection{Line and continuum luminosities} \label{sec:dis:line_cont}

\begin{figure*}[htb!]
    \centering
	\includegraphics[width=0.99\textwidth]{ 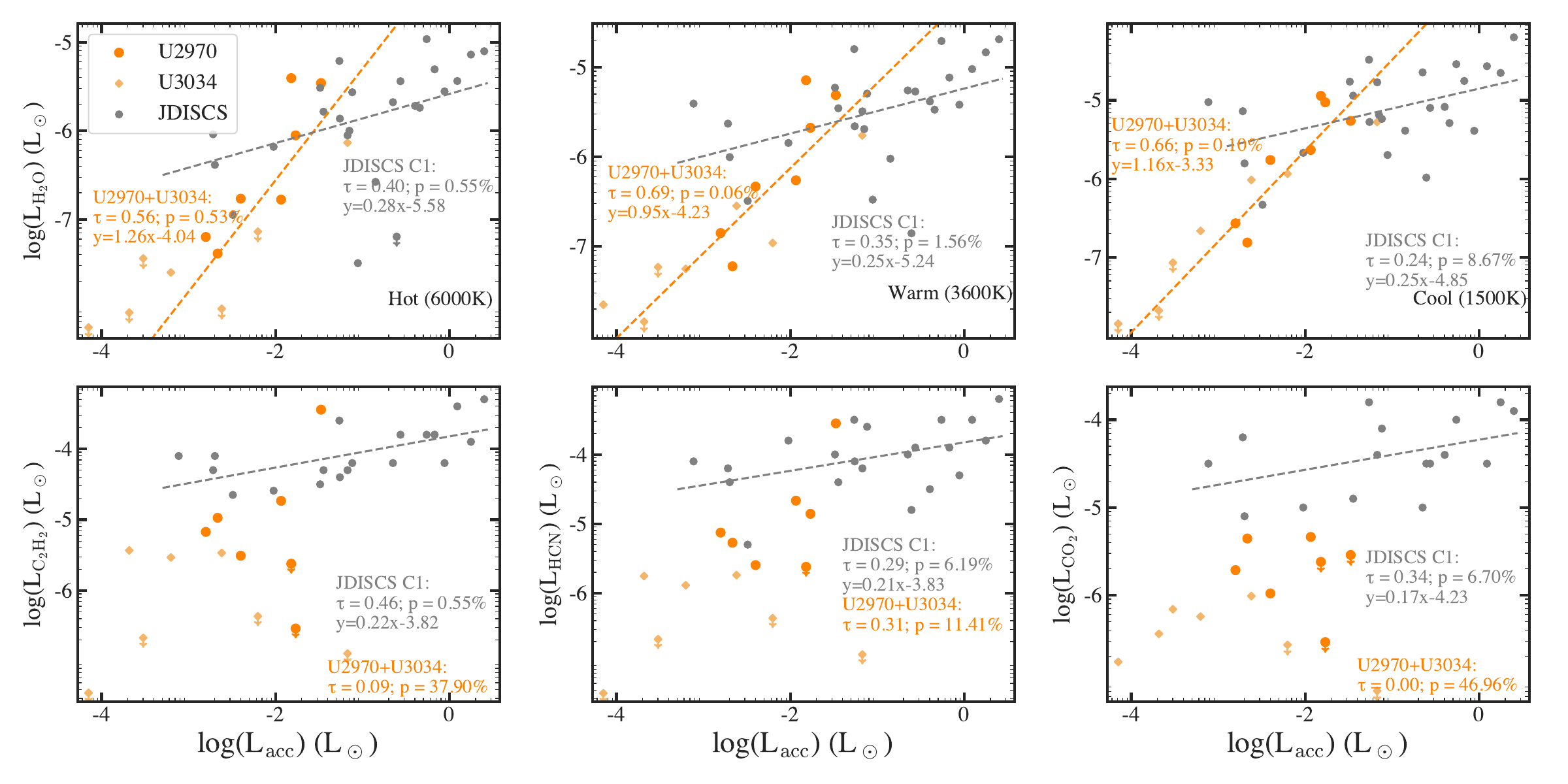}
    \caption{Comparison of molecular line luminosities and accretion luminosity. The water line luminosities are based on the hot, warm and cool single lines defined in \cite{Banzatti25}, while the C-bearing molecules are from the best-fit model within 12$-$16\,$\mu$m following \citet{Arulanantham25} and Raul et al. in prep. The Kendall $\tau$ and $p$ values are listed. The fitted correlation is overplotted if the correlation is significant with $p\lesssim5{\%}$.  
    For young disk sample, the correlation are similar for each molecule. For the older USco sample, the correlations between water lines and $\log$L$_{\rm acc}$ is stronger, while no correlation is found between C-bearing molecular luminosity and $\log$L$_{\rm acc}$. 
    }
    \label{fig:fluxes_Lacc}
\end{figure*}


In young disks, MIR emission line luminosities are known to correlate more tightly with $L_{\rm acc}$ than with $L_*$ \citep[e.g.,][]{Banzatti20}. This suggests that these lines arise from the FUV-heated surface of the disk \citep[e.g.,][]{Adamkovic16,Najita17}, with the FUV originating mostly from the accretion shock \citep{Gullbring00}.  On the other hand, the dust  continuum emission is heated by stellar optical photons incident on the disk \citep[e.g.,][]{Dullemond07}, and is hence better correlated with $L_*$. Here, we examine these correlations for the older samples to infer the heating mechanisms. As the focus is on molecular emissions, only MR disks will be considered, i.e. 7 disks from U2970, 7 disks from U3034, and all 25 young disks from JDISCS C1. Because \citet{Arulanantham25} does not provide the line luminosities of single water lines, we apply the same method described in Section~\ref{sec:res:mollinefluxes} on the published continuum-subtracted spectra to estimate them.

\begin{figure*}[htb!]
    \centering
	\includegraphics[width=0.99\textwidth]{ 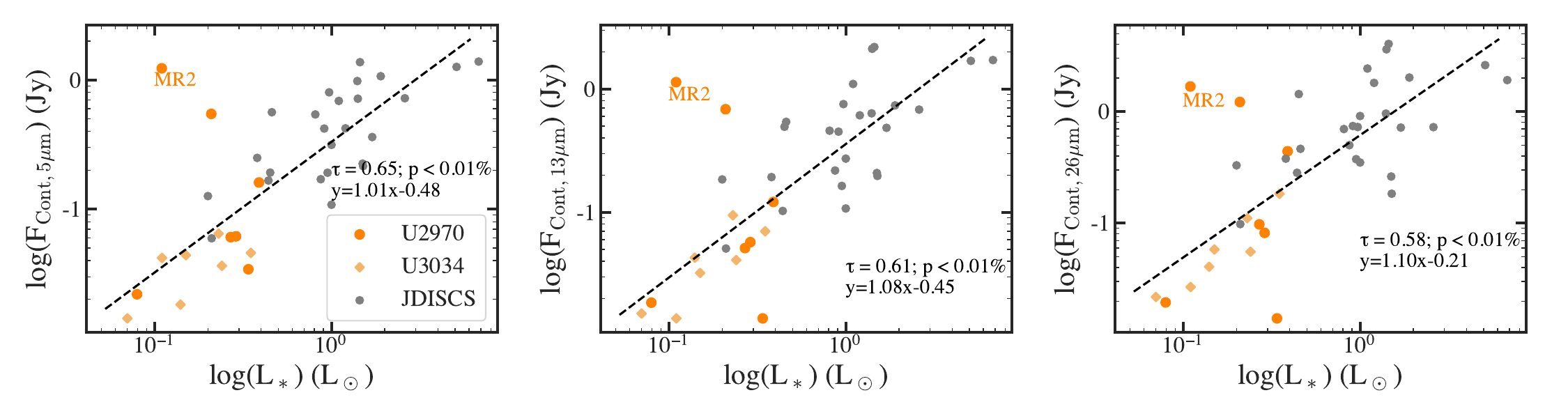}
    \caption{Comparison between the continuum fluxes scaled to 150\,pc vs. stellar luminosities $L_*$ among samples. As marked in plots, MR2 is an edge-on disk and $L_*$ of it is highly uncertain. Significant correlations can be seen between continuum fluxes and the $L_*$ for each sample, and the slopes do not change among samples (ages) and wavelengths (all the differences within 1\,$\sigma$).
    }
    \label{fig:Lacc:cont}
\end{figure*}

Fig.~\ref{fig:fluxes_Lacc} shows line luminosities versus $L_{\rm acc}$ while Fig~\ref{fig:Lacc:cont} gives continuum fluxes (at IR wavelengths) scaled to 150\,pc versus $L_*$.  To test whether there is a correlation between the two quantities, we apply the Kendall $\tau$ test as implemented in \texttt{pymccorrelation}, which takes upper limits into account \citep[see e.g.][for  applications]{2020ApJ...893..149P,2023ApJ...953..183P}. 
The correlation coefficient $\tau$ and the probability $p$ of no correlation are shown in the figures: $p<5$\% indicate a correlation while $\sim 5-10$\% only a marginal one.
When a correlation is found, we use the \texttt{GenericLikelihoodModel} in \texttt{statsmodels.base.model} \citep{SeaboldPerktold2010statsmodels}, which also accounts for upper limits, to find the best fit relation and plot it in the figures.

We find that the correlations between the continuum fluxes and $L_*$ are the same for the young and old samples, while those between the molecular line luminosities and $L_{\rm acc}$ are different. Specifically, water line luminosities show steeper correlations with $L_{\rm acc}$, while C-bearing molecule luminosities do not correlate with $L_{\rm acc}$ in the old sample. Moreover, the hot water line shows a slightly steeper correlation than the warm and cool ones both in the young and old sample (upper three panels of Fig.~\ref{fig:fluxes_Lacc}), in line with previous findings in young disks \citep[][]{Banzatti25}. 
This also agrees with theoretical expectations, in that the hot water is more sensitive to accretional heating and FUV irradiation \citep[e.g.,][]{Glassgold09,Woitke18}, or the hot water reservoir increases at a more rapid pace than warm and cool water with increased $L_{\rm acc}$ \citep[e.g.,][]{Calahan26}. 

Generally, at a given $L_{\rm acc}$, the molecular line luminosity is lower in the old disks than the extrapolated luminosities from the young disks. 
This could mean  that the emitting molecular gas is either lower in mass, and/or emitting area, and/or temperature. Our best fit models show that molecular lines are likely optically thick (compare Table~2 with table in Appendix~\ref{app:sec:tau}), while the gas temperature is lower (see Table~\ref{tab:fitting} and Appendix~\ref{app:temp}), suggesting that the lower line luminosity is primarily driven by the reduced gas temperature \footnote{Differences between emitting areas can be another cause, but they are less well constrained.}. The finding of cooler gas also agrees with the slightly higher detection rate of CO$_2$ in the old sample (see Section~\ref{sec:dis:linefluxes:detectionrates} and Raul et al. in prep.).

\subsubsection{Water line ratios}
\begin{figure}[htb!]
    \centering
	\includegraphics[width=0.5\textwidth]{ 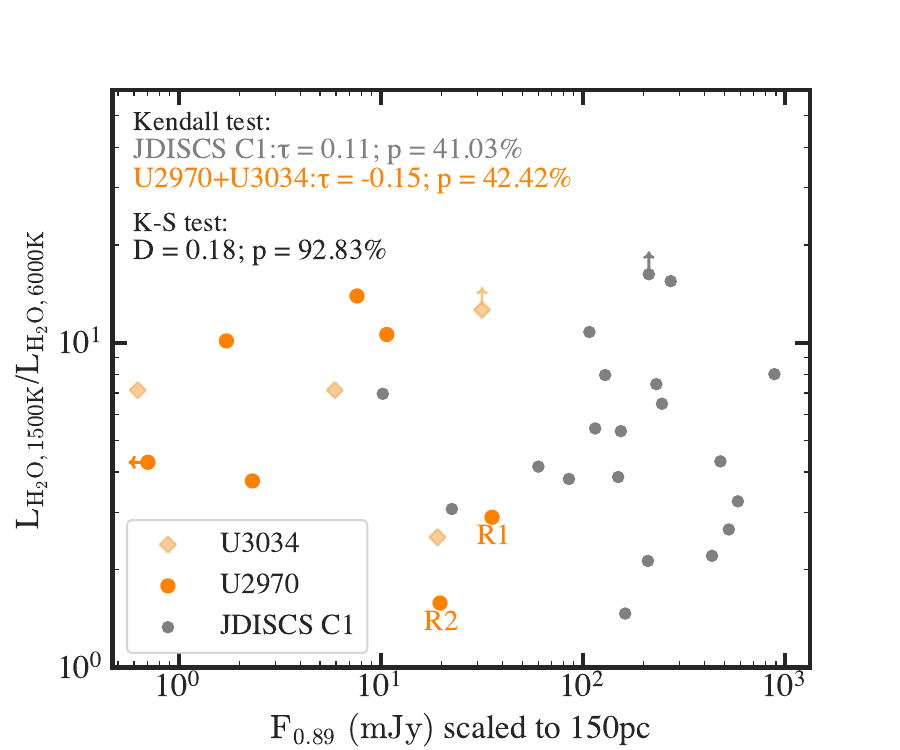}
    \caption{Line ratios of cool-to-hot water lines vs scaled millimeter fluxes. Substructures are not indicated as the ALMA spatial resolution varies among the different samples. There is no trend in the water line ratios with millimeter flux or age. 
    }
    \label{fig:WaterRatio}
\end{figure}

Isolated H$_2$O lines at 17.3 and 23.8 micron have upper energy levels of $\sim$6000\,K and $\sim$1500\,K, hence trace hotter and cooler water emission, respectively \citep{Banzatti25}. 
Although these lines strongly correlate with $L_{\rm acc}$, hence are temperature-sensitive, and likely optically thick (see Appendix~\ref{app:sec:tau}), it has been suggested that their ratio is a proxy for the mass of cool vs hot water in the observable disk surface \citep[e.g.,][]{Banzatti23,Vlasblom25}. Under the additional assumption that inward-drifting icy pebbles crossing the water snowline enhance the cool water reservoir, higher cool-to-hot water ratios have been linked to a higher pebble flux \citep[e.g.,][]{Romero24,Krijt25}. In pebble-accretion models, this inward pebble flux is a critical quantity as it directly influences the sizes and numbers of planets forming in the disk \citep[e.g.,][]{Lambrechts19}. 

It is well established that millimeter continuum fluxes, which to first order trace the amount of pebbles in the outer disk, correlate with dust disk sizes \citep[e.g.,][]{Tripathi17,Hendler20,Pinilla25}. Because dust disk sizes are shaped by radial drift and by the presence of dust traps \citep[revealed as dust substructures, e.g.,][]{Kurtovic25}, one would therefore expect that, for disks around stars of similar spectral type, larger (brighter) disks will have lower inward pebble mass fluxes, i.e. reduced inner water emission, particularly in the cool component. A trend is seen in the IR data \citep{Banzatti20,Banzatti23,Romero24}, though the scatter is large, possibly due to specific radial locations and depths of disk gaps \citep{Banzatti25,Gasman25,Krijt25}.
In addition, as disks age, their inward pebble mass flux should decline due to the overall depletion of the outer pebble reservoir \citep[e.g.,][]{Birnstiel10,Pinilla12,Kalyaan21,Kalyaan23,Mah23,Vlasblom25}.

Figure~\ref{fig:WaterRatio} compares the cool-to-hot water ratio for the young and old samples, plotted against the mm flux (F$_{\rm 0.89\,mm}$) scaled to 150\,pc. 
Across three orders of magnitude in millimeter flux (pebble mass), the cool-to-hot water ratio remains between $\sim$1 and 10 with no trend (see the large Kendall $p$ values in Figure~\ref{fig:WaterRatio}). Furthermore, the K-S test on the water flux ratios indicates that the samples of young and old disks are statistically indistinguishable. 
Notably, the two lowest orange points correspond to MR1 and MR2, which display the most water-rich spectra and have relatively high accretion luminosities.
If these two disks have a high pebble flux feeding the inner disk, the cool-to-hot H$_2$O line ratio may not be a reliable tracer of pebble mass flux as variations in the hot component can obscure any enhancement in the cool component. These results are consistent with what found in the IC~348 region, where no trend is seen between the cool-to-hot H$_2$O line ratios with age nor with the  HCN/H$_2$O and C$_2$H$_2$/H$_2$O flux ratios that, to first order, trace the C/O ratio (Carr \& Najita in prep.).  

We note that parametric models from \citet{Krijt25} would associate a cool-to-hot water ratio of 10 with a pebble mass flux of $4\times10^{-4} M_\oplus$\,yr$^{-1}$ which, if sustained for $\sim$5\,Myr (the age of USco), would correspond to 2,000\,M$_\oplus$ of dust moving into the inner disk. 
This high value is unrealistic, and disk models that properly follow dust evolution with pressure traps $-$ predicting lower pebble fluxes to the snowline $-$ are needed when comparing with observations \citep[e.g.,][]{Kalyaan23}. In addition, low disk turbulence ($\alpha$) and fragmentation velocities can suppress growth and thus radial drift to yield an approximately constant but low pebble flux over time \citep{Pinilla2025Ap&SS.370..140P}.  As such, old mm-bright disks 
may be those with low but prolonged pebble inward flux (see also Section~\ref{sec:dis:COratio} and Fig.~\ref{fig:ave:all}).

\subsection{Disk cavities/gaps and molecular emissions \label{sec:dis:cav}}
A recent study by \cite{Mallaney26} examined 12 relatively young ($\lesssim$5 Myr) disks with inner cavities observed through  various JWST Cycle~1 programs. Among them, 10 were spatially resolved in millimeter continuum with ALMA, and 2 were identified from a positive IR spectral index ($n_{13-26}$). By comparing the line luminosities of cavity disks with those of full disks, defined as having no cavities detected\footnote{limited by ALMA resolution, with the highest resolution of $\sim$5\,au} in the young JDISCS C1 sample, \cite{Mallaney26} classified cavity disks as molecular rich (CMR) or molecular poor (CMP) based on whether their hot and warm H$_2$O line luminosities (6000\,K and 3600\,K) are comparable to those of full disks. We emphasize that this CMP/CMR scheme differs from our MP/MR classification: we define MP disks as those with no detections of any of the main molecules, independent of whether a cavity is present.

In the combined USco sample, both IR indices and millimeter cavity constraints are available \citep{Vioque25,Carpenter25,Pinilla25}. We therefore classify disks as cavity or full using their millimeter continuum images \citep{Vioque25,Carpenter25}, defining cavity disks as those with emission peaks offset from the disk center. For these disks, we adopt the radius of the first emission peak as the cavity radius, $R_\mathrm{cav}$ \citep{Pinilla25}, following \cite{Mallaney26}. Disks that are unresolved at the available resolution are treated as full disks. 

Applying the CMP/CMR definition, the JDISCS~C1 sample would include two additional CMP disks (GM~Aur and RY~Lup; both show mm cavities and weak water lines). For the combined USco sample, the classification remains unchanged (all CMP are MP) except for J1622$-$2511 and J1620$-$2442 (Raul et al., in prep.).  They both have a mm cavity, and the IR spectrum of J1622$-$2511 is C-rich while for J1620$-$2442 only CO$_2$ is detected. Neither of them have water detection, hence are classified as CMP according to \citet{Mallaney26}.

\begin{figure}[htb!]
    \centering
	\includegraphics[width=0.5\textwidth]{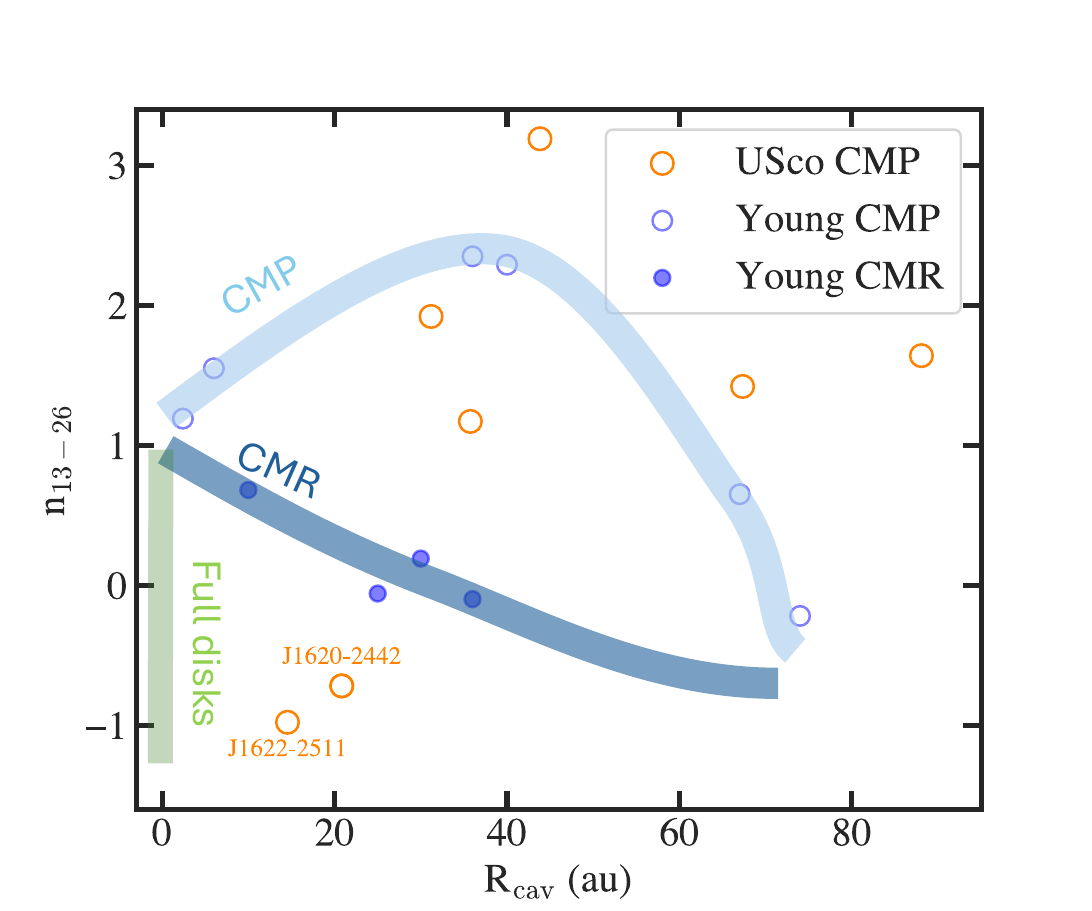}
    \caption{Our USco disks overplotted on Fig.~9 of \cite{Mallaney26}. Filled (open) circles denote CMR (CMP) disks. The shaded curves show trends identified for young  disks. In contrast to young disks, which shows a bifurcation in molecular emission depending on IR index, old USco cavity disks are all CMP (orange open circles).
    }
    \label{fig:cavity}
\end{figure}

Figure~\ref{fig:cavity} compares our older USco sample to the younger sample of \cite{Mallaney26}, who reported a clear bifurcation in IR index between CMR and CMP disks. In the young sample (blue points), disks with similar mm-cavity sizes but higher IR indices tend to be CMP (light-blue trend in Fig.~\ref{fig:cavity}), suggesting that cavities more strongly depleted in $\mu$m-sized grains have weaker molecular line emission \citep{Mallaney26}. 

In contrast, the older USco sample does not follow this CMR/CMP bifurcation: all cavity disks are CMP regardless of IR index. Together with the lower overall detection rates of common molecular lines (see Section~\ref{sec:dis:linefluxes:detectionrates}), this points to a more evolved population with generally weaker molecular emission. 

\subsection{Inner disk chemistry evolution \label{sec:dis:COratio}}
\begin{figure*}[htb!]
    \centering
	\includegraphics[width=0.99\textwidth]{ 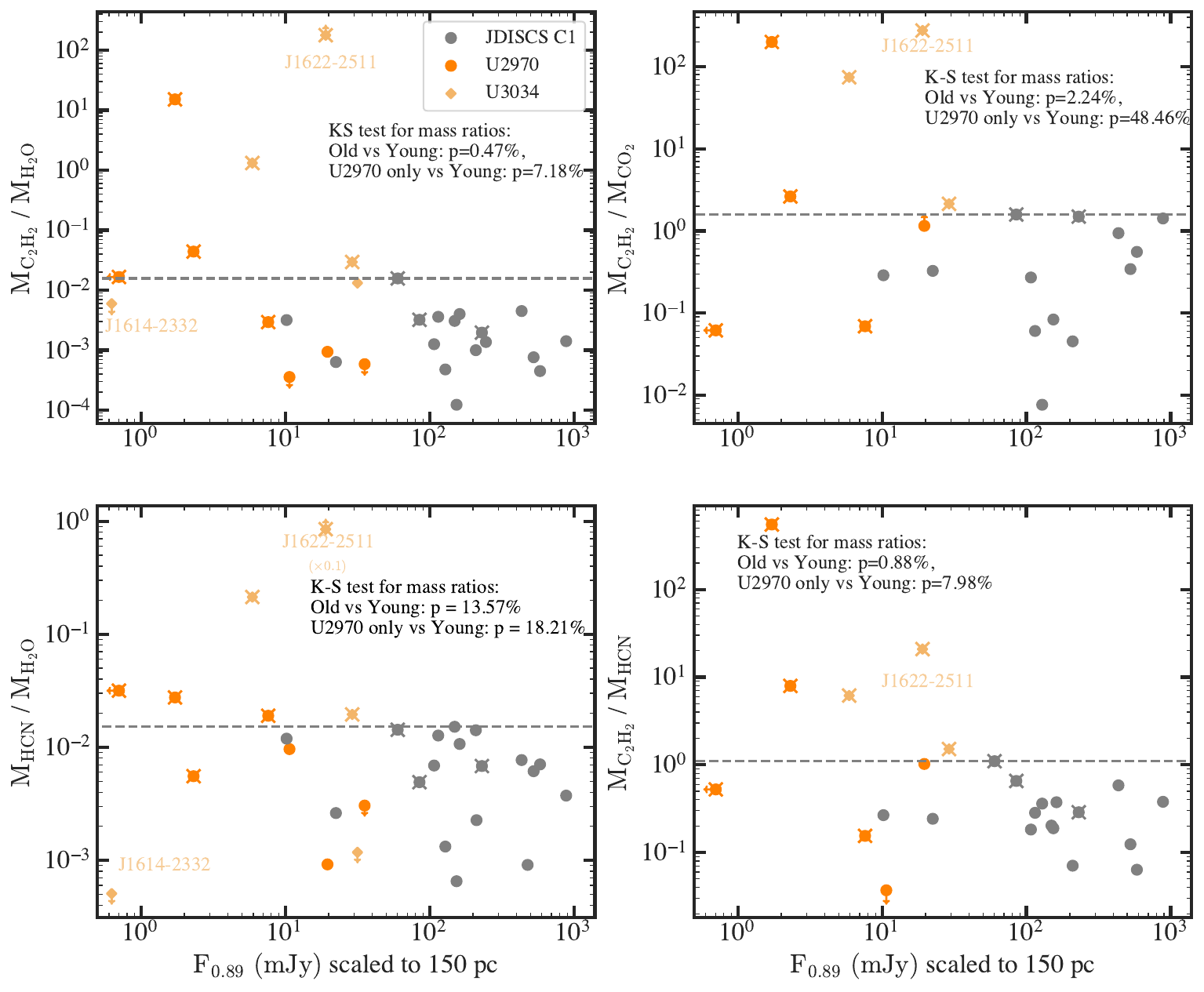}
    \caption{Observable mass ratios, expected to increase with higher gas C/O ratio,  plotted against  the 0.89\,mm flux (scaled to 150\,pc). M$_{\rm H_2O}$ is the sum of hot and warm components. Disks with a detection of C$_4$H$_2$ (hinting at elevated C/O ratios) are marked with crosses. J1622-2511 is the only cavity old disk of in this plot, while J1614-2332 shows hints of late stage infall \citep[e.g.,][]{Agurto-Gangas25,Vioque25}. Because the M$_{\rm HCN}$/M$_{\rm H_2O}$ for J1622-2511 is extremely high, we multiply the value by 0.1 to show in the plot.\\
    A gray dashed line marks the highest JDISCS C1 ratio in each panel. Around half of the old disks show higher ratios, especially those with low millimeter fluxes.
    }
    \label{fig:RatioMmol_vs_F89}
\end{figure*}

\subsubsection{Observable molecular mass ratios and average spectra \label{sec:dis:COratio:emitmass}}
The best-fit LTE models to the MIRI spectra provide  observable masses for several C- and O-bearing molecules for U2970 (this work), JDISCS~C1 (\citealt{Arulanantham25}), and U3034 (Raul et al., in prep.). The fitting methods are similar in the three studies for other molecules but differ for H$_2$O. JDISCS~C1 adopts a single H$_2$O component encompassing both hot and warm emission over a wavelength range similar to ours. In contrast, U3034 uses three components: the hot and warm components are fit in the same way as in our analysis, while the cool component is fit only at $\lambda > 18\,\mu$m.
For consistency, we define the H$_2$O observable mass for U2970 and U3034 as the sum of the hot and warm components, and compare these to the single H$_2$O emitting mass reported for the JDISCS~C1 sample.

Figure~\ref{fig:RatioMmol_vs_F89} compares observable mass ratios\footnote{We note that the upper limits of U3034 sample are recalculated for consistent mass ratios, see Section~\ref{sec:LTEfitting} for more details.} of selected C- and O-bearing species among the three samples, with each of the ratios expected to increase with higher gas C/O ratio \citep{Najita11,Kanwar26,Arabhavi26}.
Although the two samples are statistically indistinguishable, roughly half of the USco disks exhibit ratios higher than the maximum observed in JDISCS~C1 (gray dashed line), hinting at elevated inner gas C/O ratio in these systems.
This result persists when restricting the comparison to the U2970 subsample, whose spectral-type distribution  matches the JDISCS~C1 sample. Notably, the U2970 disks with higher ratios also tend to have lower millimeter fluxes and be more compact (F$_{0.89}<5$\,mJy, Table~\ref{tab:para}).

In the upper panels of Fig.~\ref{fig:ave:all}, we further compare the average spectra for the U2970 mm-bright (F$_{0.89}>10$\,mJy) and mm-faint disks (F$_{0.89}<5$\,mJy) against those of the young JDISCS~C1 sample and very-low mass stars (VLMS, $M_*<0.2M_\odot$) from \citet{Grant25}. The rationale for the latter comparison is that VLMS disks have long been known to exhibit strong C$_2$H$_2$ emission \citep{Pascucci09,Pascucci2013}, and JWST spectroscopy has now uncovered many hydrocarbons \citep[e.g.,][]{Tabone23,Arabhavi25,Long25}, confirming earlier suggestions of elevated inner gas C/O ratios (C/O$>$1). 
Fig.~\ref{fig:ave:all} illustrates that the average spectrum of the mm-bright U2970 disks (F$_{0.89}>10$\,mJy, panel b) is water-dominated and remarkably similar to the average spectrum of the younger JDISCS~C1 sample (panel a). In contrast, the average spectrum of the mm-faint U2970 disks (F$_{0.89}<5$\,mJy, panel c) is significantly more water poor, with the strongest emission arising from C$_2$H$_2$, and with additional detections of the C-bearing molecules HC$_3$N and C$_4$H$_2$. Although these features suggest higher C/O ratio, we note that this average spectrum is distinct from the average VLSM spectrum where optically thick C$_2$H$_2$ emission is present and even rarer C-bearing molecules are detected e.g., C$_6$H$_6$ (panel d).

\subsubsection{Interpretation}

\begin{figure*}[htb!]
    \centering
	\includegraphics[width=0.99\textwidth]{ 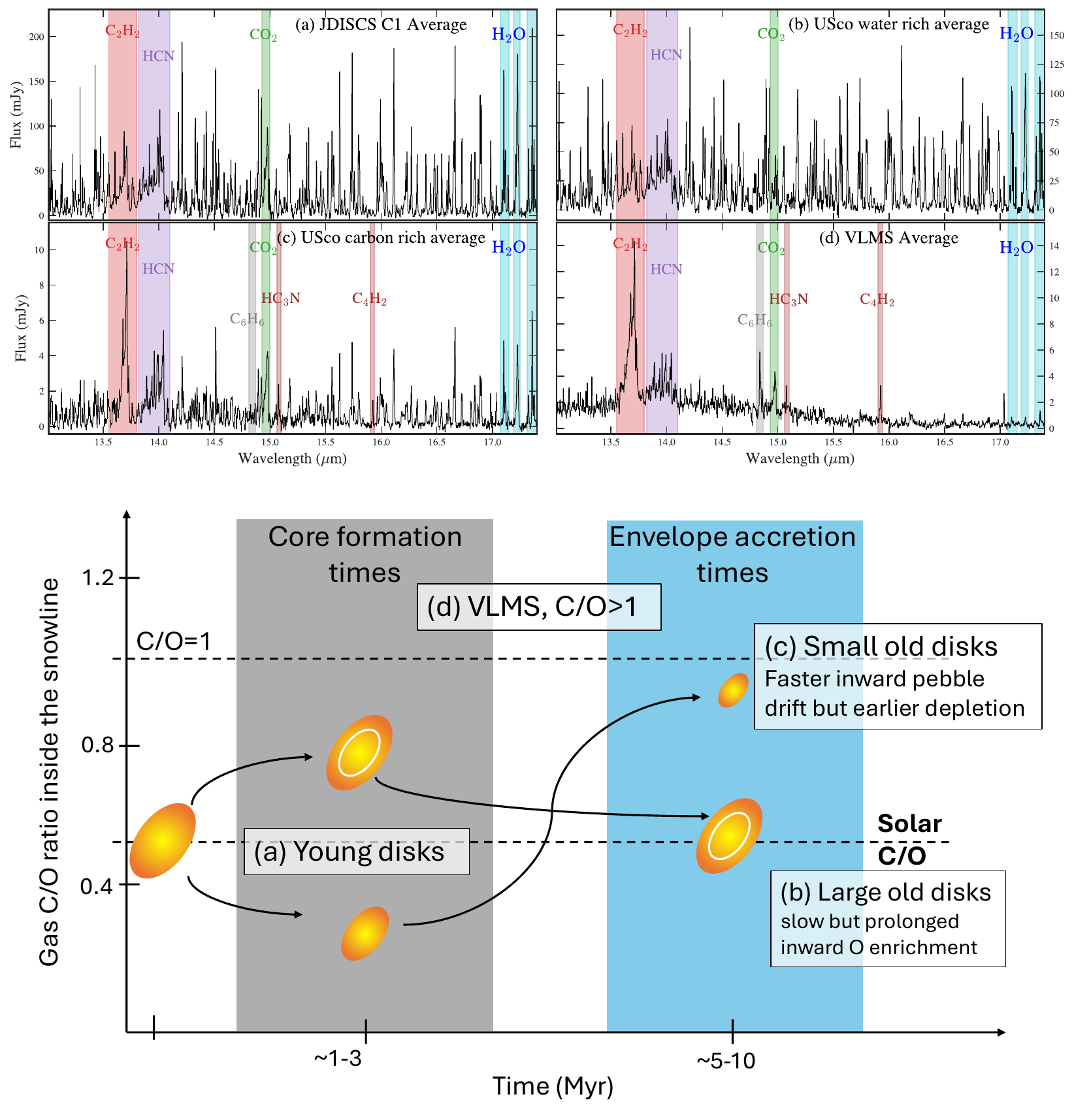}
    \caption{\textbf{Upper panel:} Comparison of average spectra from the young JDISCS~C1 sample, very low mass star (VLMS, $M_*<0.2M_\odot$, representative of C/O ratio $>1$) disks \citep{Grant25}, and the mm-bright  (MR1–MR3) and mm-faint (MR5–MR7) disks in the U2970 sample which has the same spectral type as JDISCS~C1. The mm-bright USco average spectrum closely resembles the average JDISCS~C1 spectrum. In contrast, the mm-faint average USco spectrum shows less water and stronger emission from C-bearing molecules but not at the level of the VLMS average spectrum. This suggests an elevated C/O ratio relative to the water-rich disks, but still below 1.\\
    \textbf{Lower panel:} Sketch of the proposed scenario for explaining the behavior, with (a) to (d) corresponding to the four subplots in the upper panel. }
    \label{fig:ave:all}
\end{figure*}

Three main scenarios have been proposed to explain elevated C/O ratios in the warm gas inside the snowline probed with infrared spectroscopy. Here we discuss these scenarios in the context of our findings.

\textbf{Optical depth scenario.} Disk thermochemical models show that ro-vibrational transitions from C-bearing species such as C$_2$H$_2$ probe deeper layers closer to the mid-plane than water infrared lines \citep[e.g.,][]{Woitke18}. 
A reduced dust opacity in the inner disk could expose deeper layers, making C-bearing lines comparatively easier to detect and thereby increasing their inferred observable masses relative to H$_2$O \citep{Arabhavi25}.
This scenario predicts that disks with reduced dust opacity will appear more carbon rich with a higher detection rate of C-bearing molecules \citep{Kanwar26}. 
This reduction in inner-disk dust opacity could arise from grain growth and dust settling \citep[resulting in weaker silicate features, and lower spectral slope, i.e. $F_{\rm 24\,\mu m}/F_{\rm 8\,\mu m}$, and overall lower mid-IR emission, e.g.,][]{Dullemond04,Jang25,Liu26} or through depletion of $\mu$m-sized grains \citep[which is difficult to quantify because inner disks are often optically thick; e.g.,][]{Woitke18}. 

The carbon-rich MR disks in the U2970 sample indeed show lower mid-IR continuum fluxes than the U2970 water-rich ones and the young sample, pointing to potential higher dust settling in these disks. However, \citet{Liu26} find no systematic differences in the spectral slopes, silicate feature strength or the inferred average dust grain sizes between the U2970 water-rich and carbon-rich subsets. For example, MR6 has a carbon-rich spectrum with detections of HC$_3$N and C$_4$H$_2$, but a strong silicate feature with the second smallest inferred average grain size ($\sim 0.37\,\mu$m) among the seven MR disks \citep{Liu26}. In addition, MR2 has a water-rich spectrum despite showing a weaker silicate feature and relatively low spectral slope ($F_{\rm 24\,\mu m}/F_{\rm 8\,\mu m} \sim 1.6$, \citet{Liu26}), suggesting it is more settled.  Moreover, the optical depth scenario predicts that as disks age, C$_2$H$_2$ and HCN detection rates should increase as disks evolve and dust opacity decreases. This prediction is inconsistent with the systematically lower (for all USco disks) or similar (for MR disks only)
detection rates of these species in USco with respect to the young JDISCS~C1 sample (Sect.~\ref{sec:dis:linefluxes:detectionrates}). We therefore argue that a decrease in inner-disk dust optical depth alone is unlikely to be the dominant driver of the chemical evolution from the young to the old disks. 


\textbf{Stellar radiation scenario:} Stellar irradiation through both bolometric heating and UV photons sets the disk temperature structure and the locations of key icelines. Mid-infrared water emission correlates with both $L_*$ and $L_{\rm acc}$, likely because its temperature is sensitivity to FUV radiation,
while C-bearing molecules like C$_2$H$_2$ do not correlate with $L_{*}$ in young systems \citep[e.g.,][]{Salyk11,Grant25} nor with $L_{\rm acc}$ in old disks (this work). 
Moreover, because lower UV irradiation increases observable C$_2$H$_2$ emission by reducing photodissociation \citep[][]{Walsh15}, while reducing the observable H$_2$O emission, it has been suggested that lower $L_*$ and $L_{\rm acc}$ can elevate the C/O ratios of the observable inner-disk gas \citep[see][for more details]{Colmenares24,Grant25}. While a suppressed external UV field can similarly yield higher C$_2$H$_2$ emissions in disks with C/O $< 1$ \citep{Calahan25}, this mechanism is unlikely to play a significant role in our USco sources, where the ambient external UV radiation is uniformly low \citep[$<10\,G_0$; e.g.,][]{Anania25}.

In our sample, the carbon-rich disks (MR5-MR7) have stellar luminosities that are indistinguishable from those of the more water-rich disks (MR1-MR3), but they exhibit lower $L_{\rm acc}$, corresponding to lower UV irradiation. This pattern is consistent with reduced water emission and increased survival of C-bearing molecules, making lower UV a possible contributor to the elevated C/O ratios inferred for some sources. An outlier in our sample is MR4, which has among the highest accretion luminosities yet shows a relatively carbon-rich spectrum, with strong HCN and detections of HC$_3$N and C$_4$H$_2$ (Fig.~\ref{fig:model:rich}; Table~\ref{tab:fitting}). Considering the one exception in a sample of only 7 MR disks,  larger samples of older disks with similar spectral types and UV observations will be needed to test this scenario.

\textbf{Pebble drift scenario:} Because water vapor is a major volatile and condenses onto dust grains beyond the water snowline \citep[i.e., $>2.5$~au for the Solar System, ][]{Bus02}, the outer disk gas is rich in carbon while the solids are oxygen-rich \citep[e.g.,][]{Oberg11}. As disks evolve and O-rich icy pebbles drift inwards, they release water vapor when crossing the water snowline, enhancing the inner disk with oxygen, hence lowering the inner gas C/O ratio \citep[e.g.,][]{Ciesla06,Booth19}. This low C/O ratio (C/O$\lesssim 0.5$) stage persists until the outer disk runs out of icy pebbles, or when pebble drift is halted by strong traps beyond the snowline \citep[e.g.,][]{Kalyaan21,Kalyaan23,Mah24}. Afterward, the expectation is that the inner disk gas will be dominated by the inward advection of C-rich gas from the outer disk, which lasts longer than pebble drift, and will enter a C-rich (C/O$> 1$) phase \citep[e.g.,][]{Mah23,Mah24,Houge25,Sellek25}. Moreover, as sketch in Fig.~\ref{fig:ave:all} bottom panel, the presence of pebble traps, manifesting as substructures in millimeter images of disks, can complicate this simple evolutionary picture. Larger disks with substructures, generally higher $F_{\rm mm}$ at older ages, might have less efficient but prolonged inward pebble drift and their inner C/O gas remain close to solar. In contrast, dust disks with inefficient pebble traps would shrink with time, have lower $F_{\rm mm}$ at older ages, and, upon having accreted the water vapor released by inward drifting pebbles, enter a higher C/O ratio phase. The MR disks in our sample broadly follow this scenario: the high-$F_{0.89}$ disks (MR1–MR3, all barely resolved in visibility fittings) are H$_2$O-rich, the low-$F_{0.89}$ disks (MR5–MR7, not resolved even with visibility fittings) are carbon-rich.  The intermediate mm flux source MR4 shows strong emission from both C-bearing species and H$_2$O.


Recent models that couple pebble drift with a comprehensive chemical-evolution framework reproduce the range of C/O ratios inferred for both the USco and the young sources (Molyarova et al., in prep.). In this chemically evolving model, outer-disk CH$_4$ and CO are converted into refractory hydrocarbons by cosmic ray-driven chemical processing \citep[i.e.,][]{Bergin23}. 
As pebbles drift inwards and reach the so called `soot' line ($\sim$300-500\,K), they enrich the inner disk with carbon at earlier times, slightly raising the young-disk C/O ratio above solar. 
At later times, the conversion of CH$_4$ into refractory hydrocarbons reduces the gas-phase carbon reservoir, limiting the C/O ratio in older disks and preventing it from exceeding unity $-$ yielding more C-rich spectra, but not the hydrocarbon-rich spectra observed in VLMS (see Molyarova et al., in prep. for more details).

In summary, while the reduced optical depth and the reduced FUV irradiation scenarios may account for some of the observed age trends, the pebble-drift scenario appears to better explain the overall disk evolution from the young to the older sample. We note that these  processes are not mutually exclusive and may act together and that our inference is based on a relatively small sample of older disks. 
Expanding the sample of old MR-rich disks will be key to identify which of the discussed mechanisms dominates.


\section{Summary \label{sec:summary}}
We present and analyze JWST/MIRI spectra of 14 disks around accreting stars in the older USco region ($\sim$5–10 Myr), spanning three orders of magnitude in 0.89 mm continuum flux (a proxy for disk dust mass and size). Combining this new USco sample (U2970) with the 10 disks in the AGEPRO USco sample (U3034), and comparing to the 25 young ($\sim$1-3 Myr) disks in the JDISCS~C1 sample, we find:

1. \textbf{Lower molecular detection rates in USco.} Roughly half of the USco disks are molecular poor, lacking detectable major molecular emissions whereas all young JDISCS~C1 sources have molecular rich spectra with water detections. 
These lower detection rates point to reduced molecular gas mass and/or cooler gas as disks evolve. Depite the consistently lower detectability of H$_2$O, C$_2$H$_2$ and HCN, the detection rate of CO$_2$, which probes lower temperatures, is similar in the old and young samples. This suggests cooler gas in the older sample, consistent with lower stellar and accretion luminosities. In addition, the higher detection rate of C$_4$H$_2$ in the USco sample hints at a possible increase in the inner gas C/O ratio with time. 


2. \textbf{Lower molecular line luminosities.} The correlations between molecular line luminosities and accretion luminosities  differ between the older USco sample and the young sample, while the correlation with the continuum and stellar luminosity are similar. Specifically, molecular line luminosities in the combined USco sample are on average weaker than in the young JDISCS~C1 sample.
Given that the lines are optically thick and the emitting areas are similar, though  not well constrained, the systematically lower temperatures in the older (molecular rich) sample point to reduced gas temperature as the primary driver of this trend.

3. \textbf{Elevated observable inner gas C/O ratio for smaller older disks.} Around half of the older disks (preferentially the more compact ones with low F$_{0.89}$ fluxes) have observable mass ratios of C- to O-bearing molecules that are higher than the maximum observed in the young sample. This hints at elevated gas C/O ratios in their inner disks. However, their spectra are not hydrocarbon dominated as those around many very low-mass stars ($M_*<0.2\,M_\odot$), indicating that inner disk C/O ratios of most of the older disks remain below unity. 

The higher fraction of molecular poor disks, systematically weaker line emission, and elevated inner-disk C/O ratios in the older USco sample provide  evidence for  evolution in the observable inner disk molecular layers. These trends are broadly consistent with the pebble-drift scenario, in which the inward delivery of icy pebbles declines as disks age \citep[e.g.,][]{Kalyaan23,Mah23,Krijt25}, leading to reduced water enrichment in the inner disk compared to younger systems, especially in compact ones (see Fig.~\ref{fig:ave:all} for an illustration). More broadly, this evolution implies that planetary atmospheric compositions cannot be mapped uniquely to formation locations: elevated C/O ratios may not arise only beyond the H$_2$O snowline, but can also emerge within the inner disk at later evolutionary stages. In addition, early inward transport of carbon-rich, refractory organics suggests that solids in the region between the H$_2$O snowline and the ``soot line'' ($T\sim$300–500 K) may become enriched in carbon and complex organic material. Planets that assemble their cores in this zone could therefore form as ``soot planets'' \citep[e.g.,][]{Lin25,Li26}, with interiors and envelopes, if accreted at later times, rich in complex organic molecules. 

Expanding upon the well-studied young ($\sim1-3$\,Myr) protoplanetary disks \citep[e.g.,][]{Kamp23,Arulanantham25}, our sample provides the first  insights into the inner disk chemical composition and C/O ratios of older ($\gtrsim$5\,Myr) T Tauri disks. 
However, the USco association is a large and complex region consisting of several overlapping sub-groups with ages ranging from $\sim$5\,Myr to $\sim$15\,Myr \citep[e.g.,][]{Ratzenbock23}. Our current sample of 24 disks, of which 14 are molecular rich, therefore offers only a first glimpse into the transition from young to old disks. 

Unlike USco, clusters in Corona Australis are cleanly separated in space, velocity and age \citep[][]{Posch25}. Moreover, they cover the time interval over which current data hint at an evolution in the inner disk C/O ratio.  These characteristics make the Corona Australis chain of clusters the ideal laboratory for tracing the evolution of the inner gas disk and its connection to the primordial atmospheres of forming planets $-$ objectives that will be addressed by a recently approved JWST survey (GO~10058, PI: Pascucci).




\section*{Acknowledgments}
This work is based on observations made with the NASA/ESA/CSA James Webb Space Telescope. The data were obtained from the Mikulski Archive for Space Telescopes at the Space Telescope Science Institute, which is operated by the Association of Universities for Research in Astronomy, Inc., under NASA contract NAS 5-03127 for JWST. These observations are associated with JWST GO Cycle 2 program ID 2970 (PI: I. Pascucci). Support for C.X. and I.P. through this program was provided by NASA through a grant from the Space Telescope Science Institute, which is operated by the Association of Universities for Research in Astronomy, Inc., under NASA contract NAS 5-03127.
C.X. and I.P. also acknowledge partial support from the National Aeronautics and Space Administration under agreement No. 80NSSC21K0593 for the program ``Alien Earths." The results reported herein benefited from collaborations and/or information exchange within NASA’s Nexus for Exoplanet System Science (NExSS) research coordination network sponsored by NASA’s Science Mission Directorate.
E.R. and K.Z. acknowledge the support from JWST-GO-03034.001-A.
A.E. and C.M. acknowledge support by the European Union (ERC, WANDA, 101039452). Views and opinions expressed are however those of the author(s) only and do not necessarily reflect those of the European Union or the European Research Council Executive Agency. Neither the European Union nor the granting authority can be held responsible for them.
A.E. acknowledges funding from Taighde Éireann – Research Ireland under Grant number GOIPG/2023/4396 and the UCD Physics Scholarship in Research and Teaching (SIRAT).
We acknowledge the use of the Large Binocular Telescope Interferometer (LBTI) and the support from the LBTI team, specifically from Jennifer Power, Jared Carlson, Greg Taylor. The LBT is an international collaboration among institutions in the United States, Italy and Germany. LBT Corporation Members are: The University of Arizona on behalf of the Arizona Board of Regents; Istituto Nazionale di Astrofisica, Italy; LBT Beteiligungsgesellschaft, Germany, representing the Max-Planck Society, The Leibniz Institute for Astrophysics Potsdam, and Heidelberg University; The Ohio State University, representing OSU, University of Notre Dame, University of Minnesota and University of Virginia. Observations have benefited from the use of ALTA Center (alta.arcetri.inaf.it) forecasts performed with the Astro-Meso-Nh model. Initialization data of the ALTA automatic forecast system come from the General Circulation Model (HRES) of the European Centre for Medium Range Weather Forecasts.
J. M. acknowledges support from ANID -- Millennium Science Initiative Program -- Center Code NCN2024\_001
T.M. was supported by the Royal Society, award numbers URF\textbackslash R1\textbackslash 211799 and RF\textbackslash ERE\textbackslash 231082.
All of the data presented in this article were obtained from the Mikulski Archive for Space Telescopes (MAST) at the Space Telescope Science Institute. The specific observations analyzed can be accessed via \dataset[doi:10.17909/xvc4-q391]{https://doi.org/10.17909/xvc4-q391} (for the USco sample) and \dataset[doi:10.17909/hx6h-qw97]{https://doi.org/10.17909/hx6h-qw97} (for the JDISCS C1 sample).

%

\vspace{5mm}
\facilities{JWST, ALMA, LBTI}


\software{astropy \citep{2013A&A...558A..33A,2018AJ....156..123A}, iris \citep{iris}, pymccorrelation \citep{pymc2023}, statsmodels \citep{SeaboldPerktold2010statsmodels}
          }



\appendix

\section{Binaries} \label{app:binary}
We identified 4 binary candidates based on our datacubes, listed in Table~\ref{app:tab:binary}. All these four are too close with each other, and can only be separated in channel 1 of our MIRI data.  To identify the spectra for each of the binary candidates, we generate and
we apply the following procedure to fit the data.  
We note that J16153456-2242421 (MR3) is also a binary system but with a separation of $>1''$, which can be well separated spectrally and not discussed in this section. 

\begin{deluxetable}{c|cc}
\tablecaption{Newly identified binary candidates} \label{app:tab:binary}
\tablewidth{0.99\textwidth}
\tablehead{
Coordinates &  Separation($''$) & Notes \\
}
\startdata
J16141107-2305362  & 0.41 & \\ 
\hline 
J16062196-1928445  & 0.59 & \\
\hline
J16120505-2043404 & 0.43 & \\
\hline
J16153220-2010236  & 0.32 & \\
\enddata
\end{deluxetable}

\begin{figure*}[htb!]
    \centering
	\includegraphics[width=0.49\textwidth]{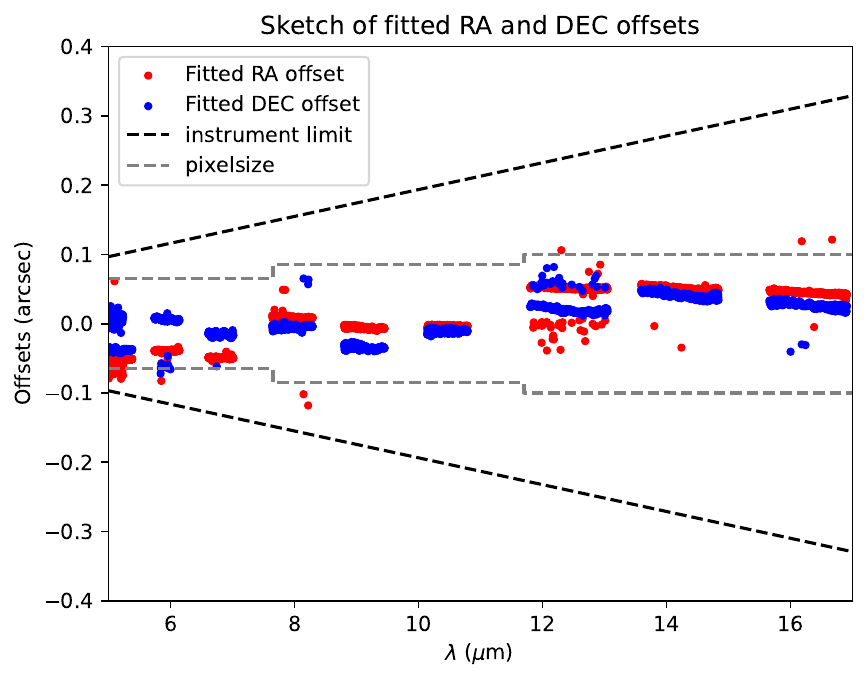}
    \caption{Fitted offsets for RA and DEC compared with MIRI resolving power and pixel sizes. 
    }
    \label{app:location}
\end{figure*}

\textbf{1. Verify consistency of peak locations across wavelengths.} Because at longer wavelengths the primary and secondary sources are not separated, it is essential to ensure their peak locations remain consistent across all wavelengths for accurate fitting. We validate this with single stars in our sample. First, we generate the point spread function (PSFs) for the JWST MIRI instrument with the \texttt{jwst\_stpsf} package. We then apply the least-square minimization code \texttt{lmfit} \cite{lmfit14} to fit the PSFs to our single star disks across different channels and bands. During the process, both the peak intensities and positions are treated as free parameters. We then estimate the uncertainty of the fitted center positions of the stars at different wavelengths. The differences of center locations are within $\sim$0.05$''$ among different bands, and are well within the instrument resolving power ($1.22 \lambda/D$, see Fig.~\ref{app:location}). 

\textbf{2. Determine the centers of binary components.} Once positional consistency is confirmed across wavelengths, we locate the centers of the primary and secondary components for our binary candidates using the Channel 1 data cubes, where the two stars are spatially separated. We apply a similar PSF-fitting method as with the single stars but fit two PSFs simultaneously to each wavelengths, allowing both peak intensities and positions to vary. The average fitted right ascension (RA) and declination (DEC) are then recorded as the locations for each component star.

\textbf{3. Estimate Flux Ratios Between Primary and Secondary Components.} With the positions of each binary component fixed, we perform PSF-fittings on the data cubes at all wavelengths with only the peak intensities as free parameters. This approach enables us to estimate the flux ratios between the primary and secondary components at each wavelength. Using the fitted flux ratios, we can decompose the total observed spectra at each wavelength to generate the individual spectra for the primary and secondary sources, and generate the spectra for each sources. 

\begin{figure*}[htb!]
    \centering
    \subfigure[J16120505]{
        \includegraphics[width=0.475\textwidth]{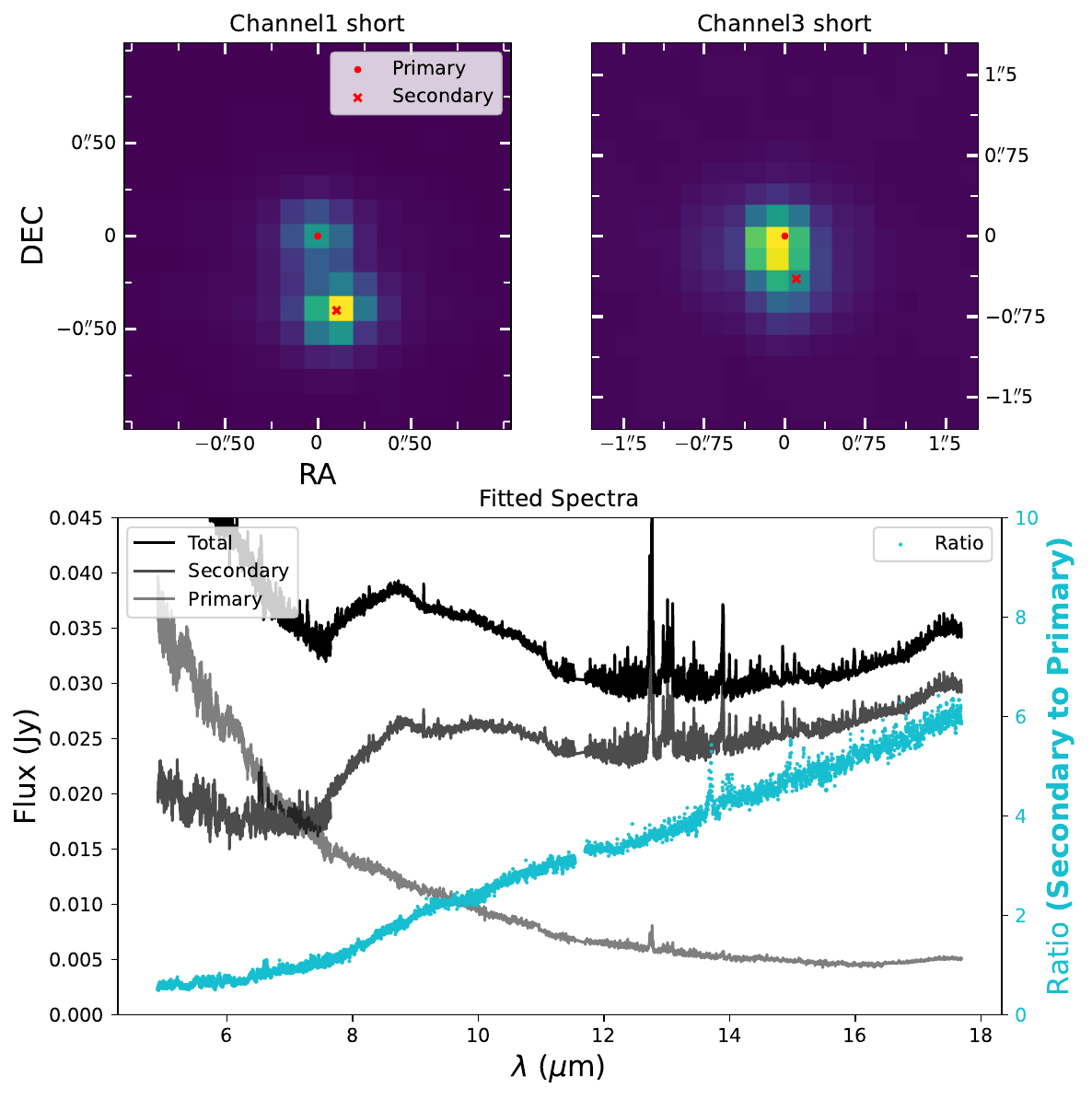}
        \label{app:Total:J16120505}
        }
    \quad
    \subfigure[J16153220]{
        \includegraphics[width=0.475\textwidth]{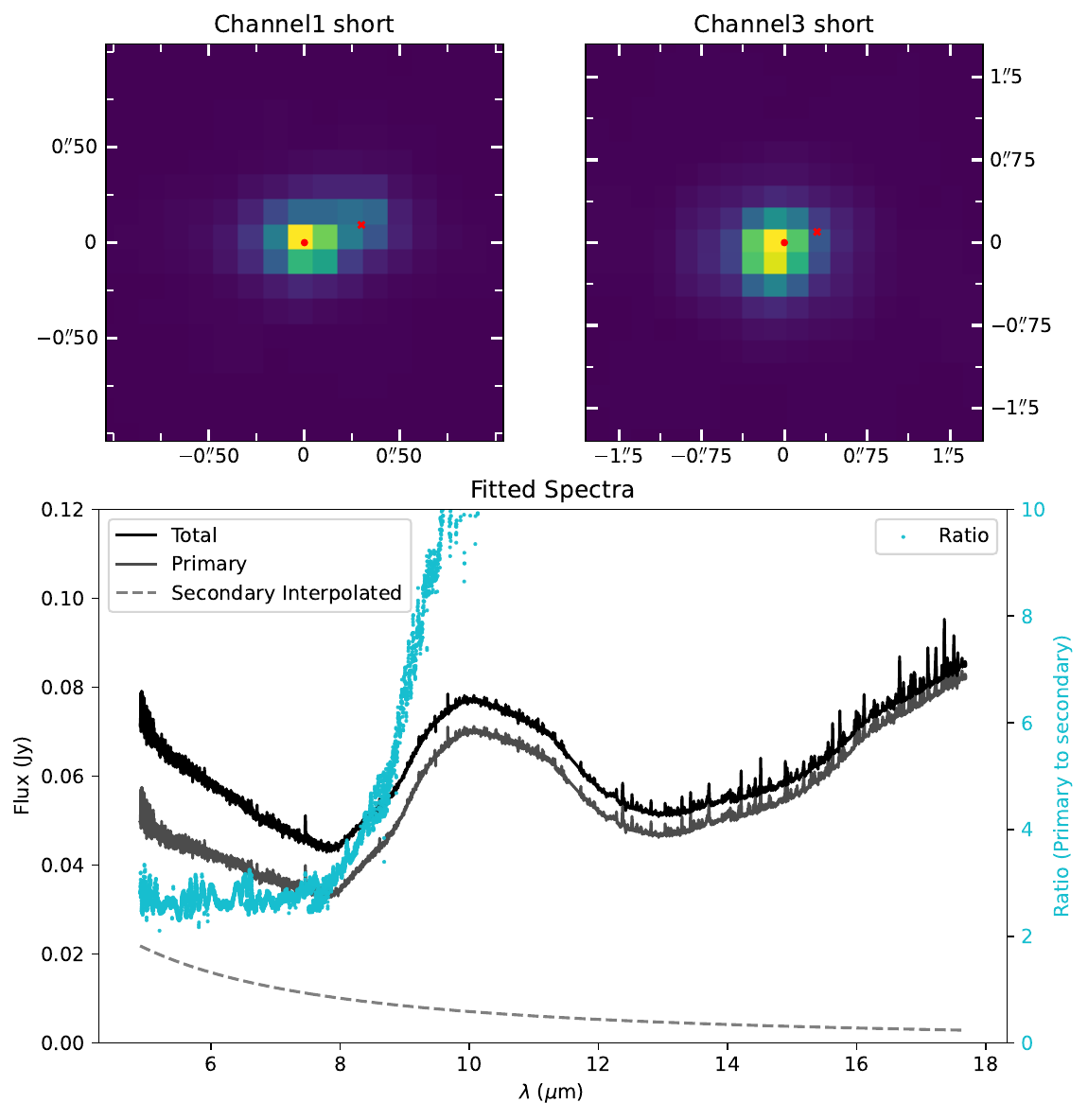}
        \label{app:Total:J16153220}
        }
    \quad
    \subfigure[J16062196]{
        \includegraphics[width=0.475\textwidth]{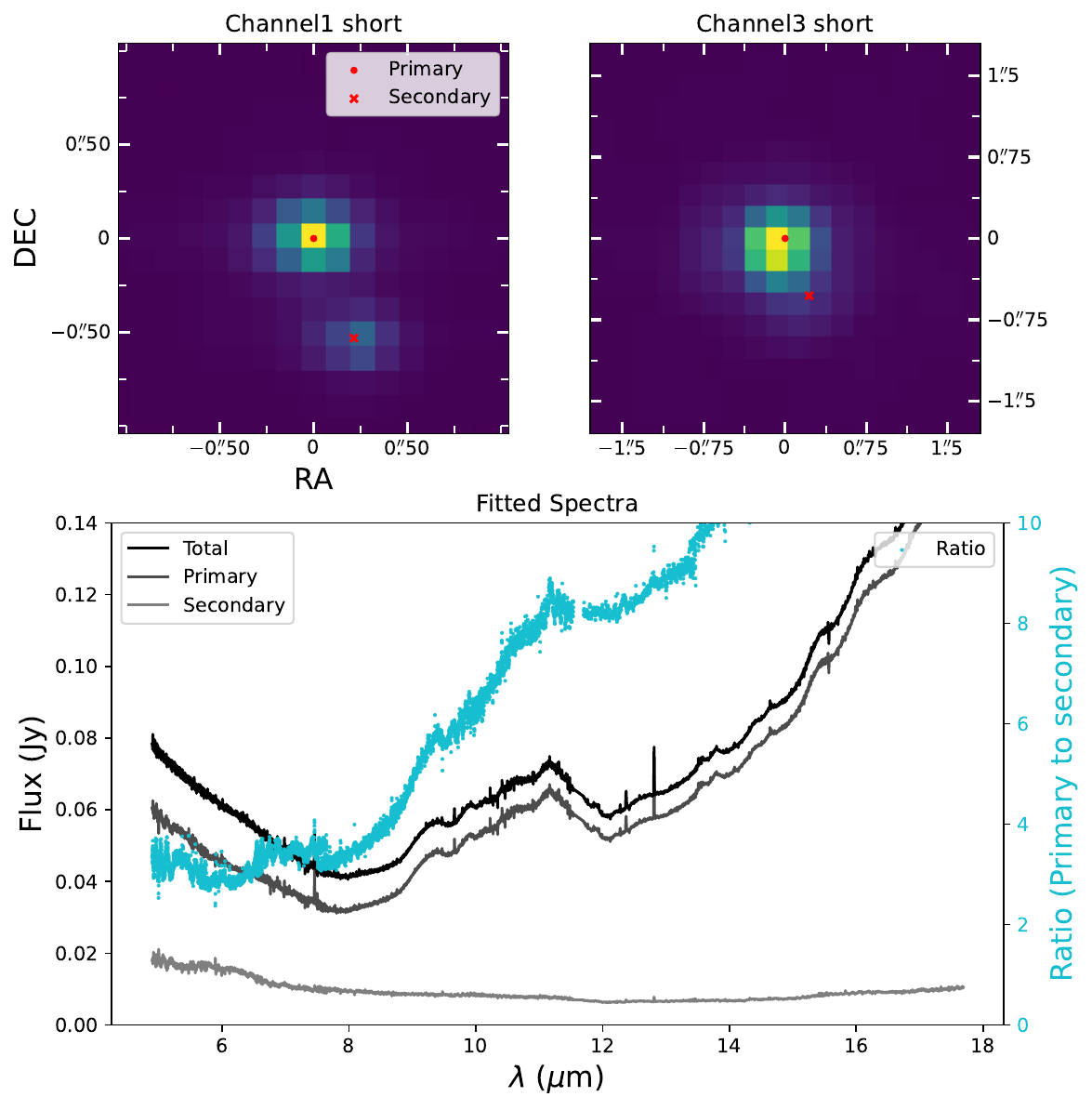}
        \label{app:Total:J16062196}
        }
    \quad
    \subfigure[J16141107]{
        \includegraphics[width=0.475\textwidth]{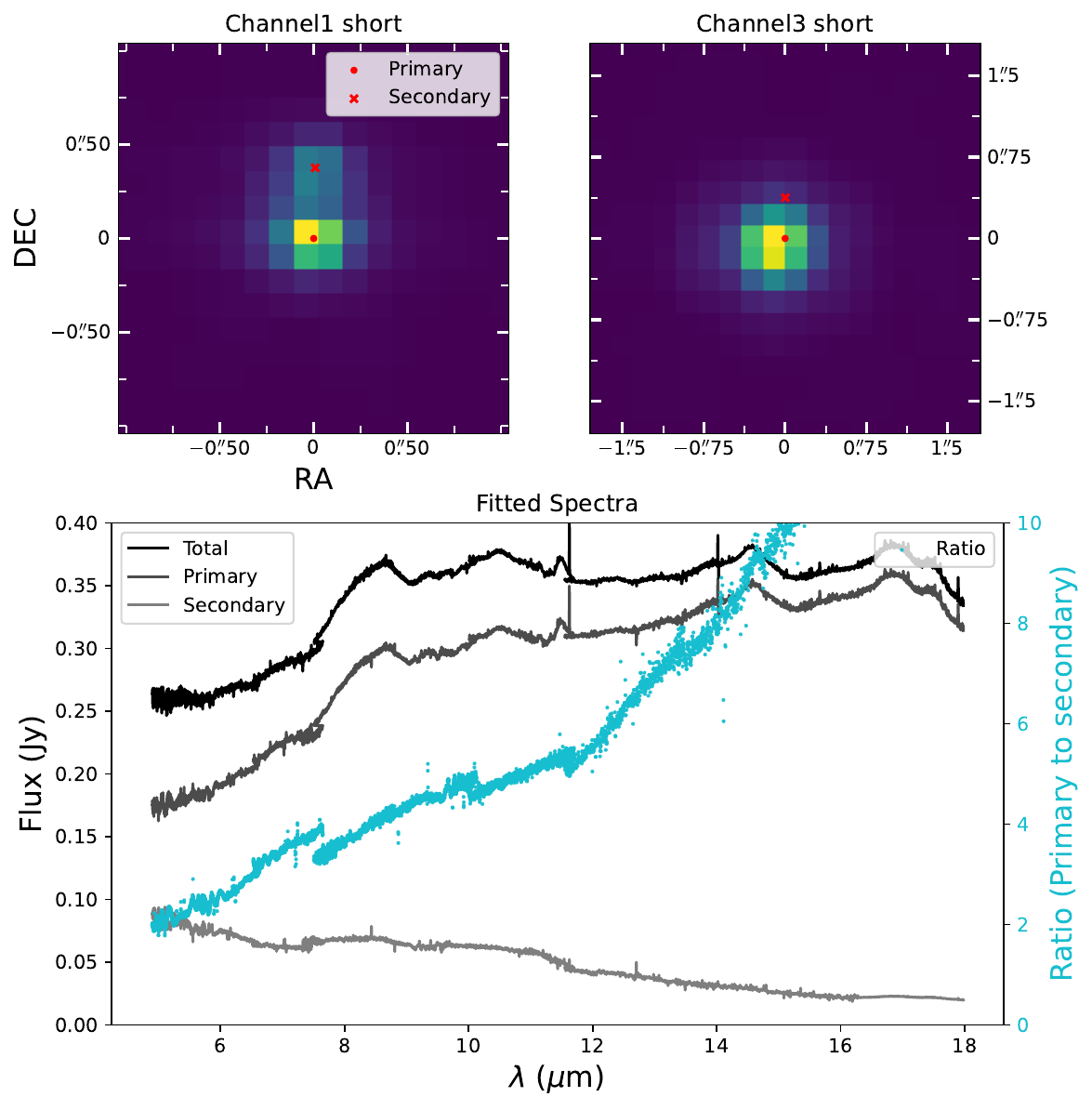}
        \label{app:Total:J16141107}
        }
    \caption{The PSF fitting results for the four newly identified binary candidates. \label{app:fig:Total}}
\end{figure*}

The PSF fitting results for all four binary candidates are shown in Fig.~\ref{app:fig:Total}. Among all the candidates, the primary sources of three targets (J16141107, J16062196, and J16153220) are much brighter than the secondary sources ($F_{p}/F_{s} > 10$). For these systems, the secondary sources will not significantly influence the spectral type classification. 
For J16120505, however, the primary star which is brighter at shorter wavelengths (e.g., channel 1, $\lambda < 7.5\mu m$) is a photosphere-like object and dimmer at longer wavelengths according to our fitting, see Fig.~\ref{app:Total:J16120505}. The spectral features (e.g., molecular line emissions) present in the primary component is a small fraction ($<10\%$) of the total spectra and are likely the residuals of the PSF fitting. After cross-checking the locations of the ALMA image and our JWST cubes, we found that all the disk emission was from the secondary object. 

\begin{figure*}[htb!]
    \centering
	\includegraphics[width=0.40\textwidth]{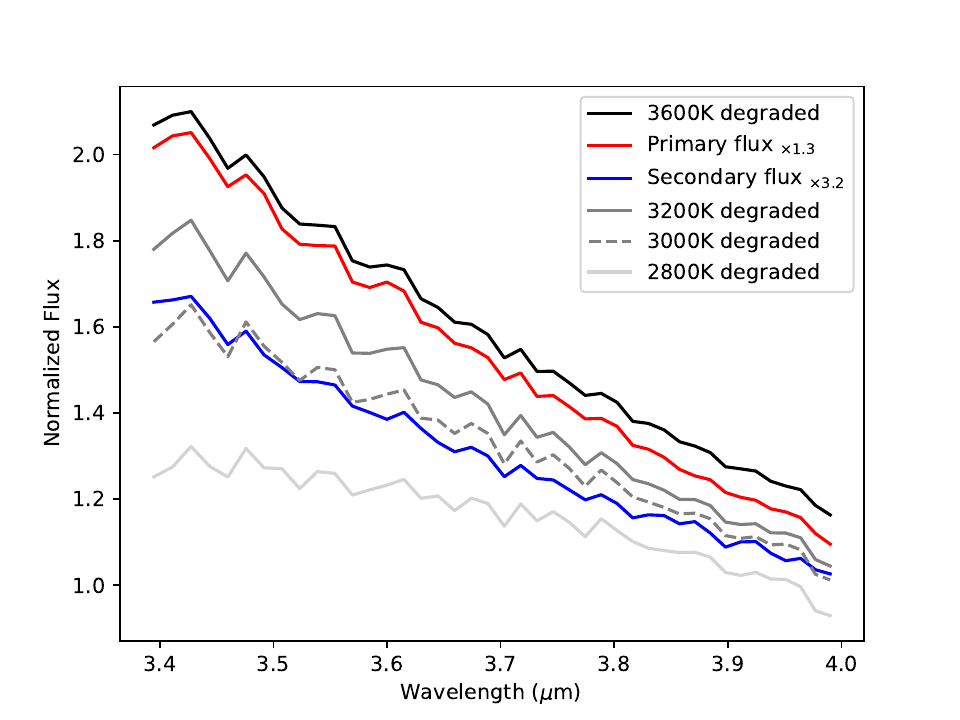}
    \caption{Spectra of the two components of J16120505 compared with the normalized photosphere spectra of different temperatures. Considering the total flux of J16120505 is consistent with a M1.5 star \citep{Fang23}, the primary component corresponds to a $\sim$3600\,K star ($\sim$M1.5), and the secondary component corresponds to a $\sim$3100\,K star ($\sim$M4). 
    }
    \label{app:fig:LBTI}
\end{figure*}

To determine the spectral types of each components for J16120505, we acquired LBTI/ALES \citep[][]{LBTI16,LBTI20,ALES20,ALES22} spectra for the source on 2025 Jun 21. 
The two components of the binary candidate are well separated, and the retrieved normalized spectra are shown in Fig.~\ref{app:fig:LBTI}. For comparison, we also plot the stellar photosphere models of 2800-3600\,K stars from BT-Settl model, convolved to the spectral resolution of LBTI/ALES. 

Based on the PSF modeling and the LBTI/ALES observations of J16120505, the primary component of it is a photosphere-like object with a stellar luminosity $\sim$3 times of the secondary component. Considering the total flux of J16120505 is consistent with a M1.5 star \citep{Fang23}, the primary component corresponds to a $\sim$3600\,K star (still $\sim$M1.5), and the secondary component corresponds to a $\sim$3100\,K star ($\sim$M4). We note that this is a very rough estimation and further higher resolution spectra in optical or NIR wavelengths will be crucial for determining the spectral types of them. 

In the separated spectra of J16141107, clear features can still be seen in both spectra. Unlike J16120505, however, the features that appear in the secondary source of J16141107 are not visible in the total (unresolved) spectrum. These features may arise from imperfections in the PSF fitting, but they could also originate from the secondary component itself.

\section{High-energy photon ionization of Neon and Argon}\label{app:sec:dis:ionization}
The Ne$^{+}$ and Ar$^{+}$ ions detected in disks can only be ionized by stellar EUV (13.6 eV $< h\nu <$ 100 eV) and/or X-ray (0.1 keV $< h\nu <$ 10 keV) photons due to their high ionization potentials. Following the literature \citep{Hollenbach09,Szulagyi12,Bajaj24}, soft/hard EUV are defined as $h\nu \lesssim$ or $\gtrsim$ 41 eV, while soft/hard X-ray are defined as 0.1 keV $\lesssim h\nu \lesssim$ 0.3 keV and $\gtrsim1$\,keV, respectively. [Ne\,{\scriptsize II}]/[Ne\,{\scriptsize III}] and [Ne\,{\scriptsize II}]/[Ar\,{\scriptsize II}] line intensity ratios can help determining whether soft/hard EUV or soft/hard X-ray stellar photons provide the dominate source of disk atmosphere ionization. 

\begin{figure*}[htb!]
    \centering
	\includegraphics[width=0.90\textwidth]{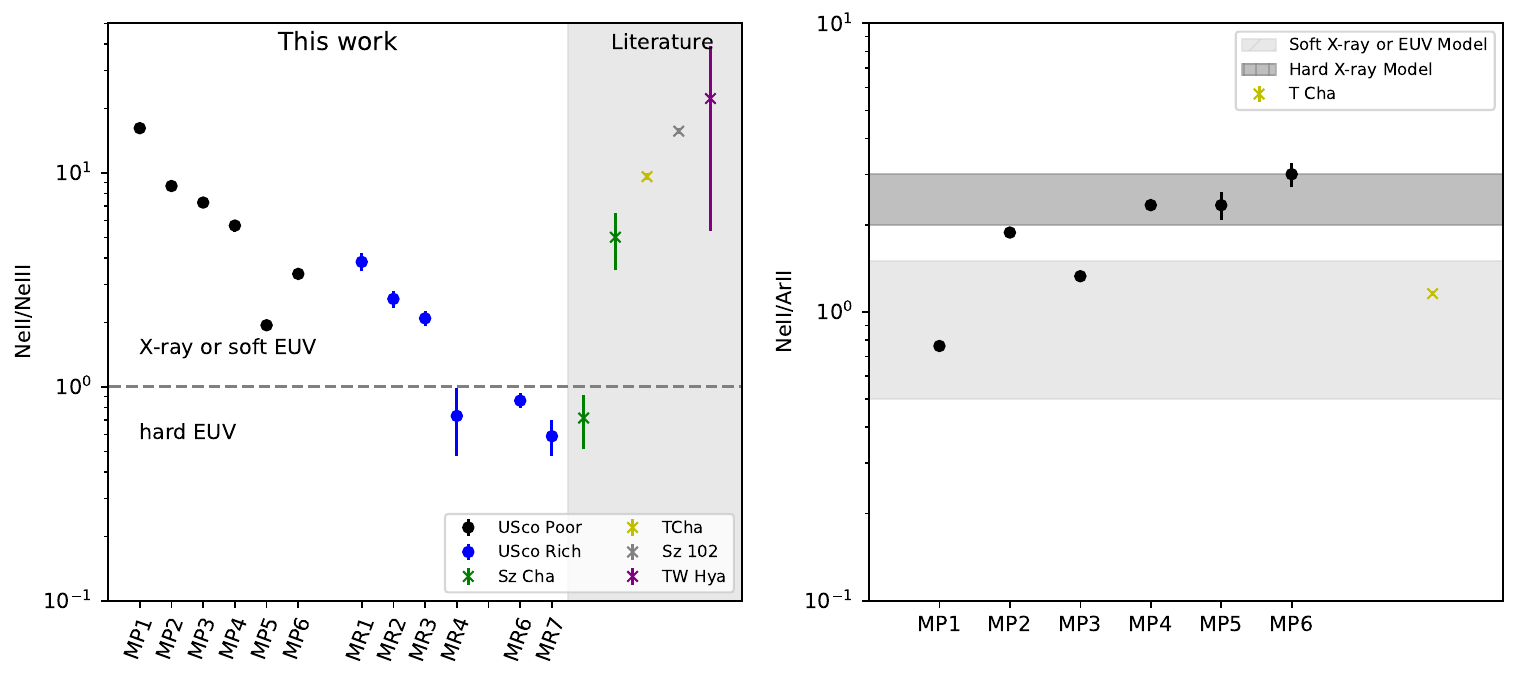}
    \caption{Ratios of ionized lines compared with models. We also show literature values for sources where both [Ne\,{\scriptsize II}] and [Ne\,{\scriptsize III}] are detected: Sz Cha \citep{Espaillat23}, T Cha \citep{Bajaj24}, Sz 102 \citep{Lahuis07}, and TW Hya \citep{Najita10}. Most line ratios point to X-ray or soft EUV ionization, only in two disks (MR6 and MR7) ionization by hard EUV photons may be prevalent. 
    }
    \label{app:fig:models}
\end{figure*}

In Fig~\ref{app:fig:models}, we plot the [Ne\,{\scriptsize II}]/[Ne\,{\scriptsize III}] for MP+MR and [Ne\,{\scriptsize II}]/[Ar\,{\scriptsize II}] line ratios for MP disks only due the contaminatio with water lines, see Sect.~\ref{sec:H2_ionic_fluxes}. On the same figures, we also overplot model predictions from \cite{Hollenbach09} and literature values of other disks detected with these ionic forbidden lines for comparison. The [Ne\,{\scriptsize II}]/[Ar\,{\scriptsize II}] model assumes solar Ne/Ar elemental abundance for comparing the two species. 

Most of the disk in our sample shows a [Ne\,{\scriptsize II}]/[Ne\,{\scriptsize III}] value larger than 1, consistent with ionization from X-ray or soft EUV photons. These values, combined with the line ratios of [Ne\,{\scriptsize II}]/[Ar\,{\scriptsize II}] and the assumption of solar Ne/Ar elemental abundance, indicate that soft X-ray is the main ionization source of MP1 and MP3, while hard X-ray is the main ionization source of MP2, MP4, MP5, and MP6. Three of the molecular rich disks (MR4, MR6 and MR7) have the NeII/NeIII ratios less than unity, indicating potential EUV ionization. EUV ionization is a rare case \citep[e.g.,][]{Szulagyi12}, and has only been found in the {\it Spitzer} observation of Sz Cha, while the JWST spectrum of the same source indicates X-ray or soft EUV ionization \citep{Espaillat23}. 

\section{Accretion luminosity measurements for different samples \label{app:Lacc}}

In this work, we use literature accretion luminosities based on the Balmer jump or optical H and He emission lines as they have been extensively calibrated \citep[e.g.,][]{Herczeg08,Manara13,Alcala17,Fang23,Manara23,Fiorellino25}. Specifically, for the U2970 sample, accretion luminosities are estimated from two Balmer lines (H$\beta$ and H$\alpha$) and two He\,{\scriptsize I} lines (5876 and 6678\,\AA) with HIRES spectra \citep[for more details, see][]{Fang23}. For JDISCS C1 and U3034, values are compiled from \citet{Manara23} and Empey et al. in prep., respectively. In both studies, accretion luminosities are derived from the UV continuum excess (encompassing the Balmer jump) measured with respect to a photospheric template and a grid of isothermal hydrogen slab models \citep[see][for details]{Manara13,Claes24}. 
In this paper and all the studies mentioned above, the conversion between the accretion luminosities and mass accretion rates \citep[which are the values reported in][]{Fang23,Arulanantham25} are calculated with the relation $\dot{M}_{acc} = \frac{L_{acc}R_*}{GM_*} ( 1- \frac{R_*}{R_{in}})^{-1}$, with the typical assumption of $R_{in} = 5R_*$ \citep{Gullbring98}.
A comparison of the two methods for 21 accreting USco sources showed an overall good agreement with a mean difference in the accretion rates of 0.38\,dex, within the quoted uncertainties of the two methods \citep[see e.g., Appendix~G and Fig~29 in][]{Fang23}. The only exception is J16064385 (MR7), which differs by $\sim$1.5\,dex and may have undergone accretion outbursts. Its photosphere-like SED, weak silicate feature, and ALMA non-detection suggest that the strong molecular emission may be driven by variable accretion \citep[e.g.,][]{Smith25}, making it an excellent target for follow-up observations.

\begin{figure*}[htb!]
    \centering
	\includegraphics[width=0.99\textwidth]{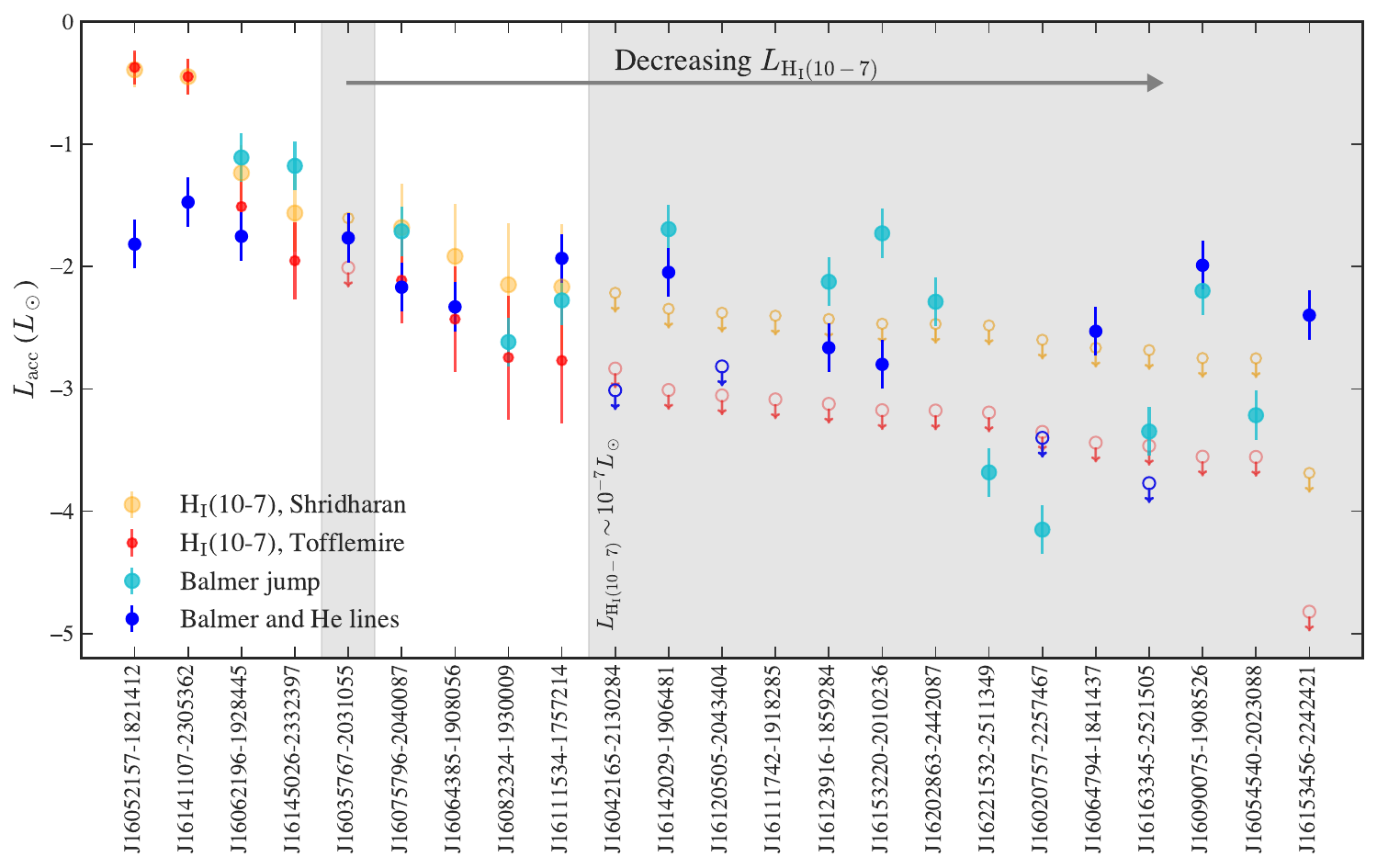}
    \caption{$L_{\rm acc}$ values for USco sources derived from different methods: i) H\,{\scriptsize I} (10$-$7) from \citet{Shridharan25} and \citet{Tofflemire25} (orange and red points), respectively; ii) Balmer and He lines from \citet{Fang23} (dark blue points); and iii) Balmer jump from Empey et al. in prep. (light blue points). Upper limits are indicated as empty symbols with an arrow pointing down. Sources are ordered from high (left) to low (right) 
    H\,{\scriptsize I} (10$-$7) luminosities, with non-detections indicated as gray regions. 
    More than half of the USco sample is not detected in the  H\,{\scriptsize I} (10$-$7) line, and for the two highest accretors  the $L_{\rm acc}$ values from the H\,{\scriptsize I} (10$-$7) lines are more than 1.5\,dex higher than those derived from optical spectra. Values from \citet{Fang23} and Empey et al. in prep. (dark and light blue points) are generally consistent with each other. 
    }
    \label{app:fig:Lacc:method}
\end{figure*}

Permitted hydrogen lines are also covered at infrared wavelengths, and {\it Spitzer} spectra have been previously used to calibrate the H\,{\scriptsize I} (7$-$6) lines against H$\alpha$, noting the need for proper removal of water emission overlapping with the (7-6) transition \citep[e.g.,][]{Rigliaco15}.
Recently, \citet{Tofflemire25} used JWST/MIRI to re-calibrate the H\,{\scriptsize I} 7$-$6 and other H\,{\scriptsize I} lines (i.e., 6$-$5, 10$-$7) using nine epochs of contemporaneous VLT/X-Shooter spectroscopy of a single source, DQ~Tau, which is a binary system with accretion bursts every orbital period. The even more recent work by \citet{Shridharan25}, also using JWST/MIRI, correlates these same H\,{\scriptsize I} transitions with non-contemporaneous literature accretion luminosities for 79 sources. The two JWST works arrive at different conclusions regarding the correlations, especially for the 7-6 transition which remains contaminated by water even at the higher resolution of MIRI. As such, we show in Fig~\ref{app:fig:Lacc:method} the comparison of the $L_{\rm acc}$ derived from the H {\scriptsize I} (10-7) transition, which is not contaminated by water, with the literature values based on optical lines or Balmer jumps. Even for this (10-7) transition, we notice a systematic difference of $\sim 1$\,dex between the two works, with \citet{Tofflemire25} giving lower $L_{\rm acc}$ for the ones with relatively low $L_{\rm H \scriptsize I}$. More importantly, more than half of the USco disks have a non detection in the H\,{\scriptsize I} (10-7) line (grey region) suggesting $L_{\rm acc}$ significantly lower, at times by more than an order of magnitude, than what measured with the Balmer jump and optical lines. While the agreement among these and the optical tracers is good for $L_{\rm H {\scriptsize I} (10-7)} \sim 10^{-6}\,L\odot$, we see that for the two highest accretors (MR1 and MR2), the values derived from the H\,{\scriptsize I} (10$-$7) lines are more than 1.5 dex higher than those from optical lines \citep{Fang23}, well above the typical accretion variability of Class~II sources \citep[e.g.,][]{Fischer2023ASPC..534..355F,Manara23,Pittman25}. This difference may be due to additional contributions to the infrared H lines, like winds and jets \citep[e.g.,][]{E2010MNRAS.406.1553E,Bajaj25}. 

Overall, our comparison suggests that the extension of the correlation between the H\,{\scriptsize I} 10-7 line and accretion luminosity inferred from the Balmer jump and optical lines in young sources should be further tested in older stars and at low $L_{\rm acc,IR}$ $\lesssim 10^{-3}$\,L$_\odot$. Thus, we argue that for the older USco sample the literature values derived from optical spectra are better tracer for the accretion luminosity for this analysis of older stars and low accretion rates objects

\section{Model comparison \label{app:model}}
In this paper, the whole USco sample consists of two separate samples, U2970 and U3034. We directly use the best-fit models from Raul et al. in prep. for U3034 and combine them with our best-fit models for U2970. The data reduction, continuum subtraction and the fitting procedures (LTE slab models) used in Raul et al. in prep. are the same as in this work, but the fitting for water is different. For U2970, to get a better constraint on water fluxes to minimize the contamination to the C-bearing molecular emission within 12$-$16\,$\mu$m, we only consider hot and warm components and fitted within 11$-$19\,$\mu$m. For U3034, on the other hand, Raul et al. in prep. fits the water emission across a broader range, and included hot, warm and cool components together. 

\begin{figure}[htb!]
    \centering
	\includegraphics[width=0.49\textwidth]{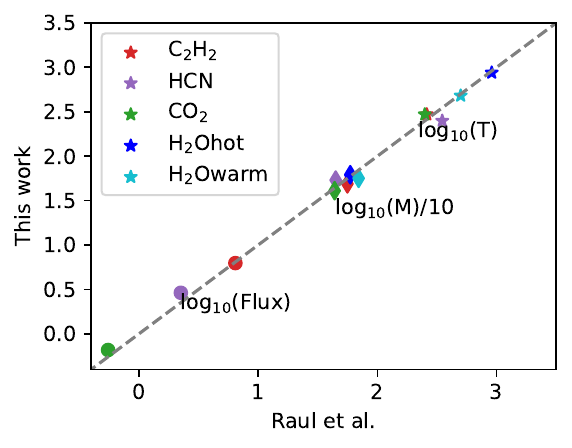}
    \caption{Comparison between the model from Raul et al. and our model. T is the fitted temperature in K,  M is the total emitting mass defined as $N\times \pi R^2$ in kg. Flux is integrated within 12-16$\mu$m based on the model, in 10$^{-15}$\,erg/(s cm2). We can see our models are consistent with the model in Raul et al. Specifically, our hot and warm water models correspond to the hot and warm water models in Raul et al. 
    }
    \label{app:fig:comparison}
\end{figure}

To ensure consistency between the U2970 and the U3034 results, we apply our fitting methods to two targets (J1614-2332 for water fittings and J1622-2511 for C-bearing molecule fittings) in U3034 and compare them with the results from Raul et al. in prep. The results are summarized in Fig.~\ref{app:fig:comparison}. For the C-bearing molecules, as we are using the models to estimate the fluxes for each molecule, we also show the fluxes derived for each model. Because column density and emitting area (radius) are highly degenerate, we only show the better constrained total emitting mass (M). Our results are consistent with those in Raul et al. in prep.. 

\section{Model temperature and emitting area comparison\label{app:temp}}
\begin{figure}[htb!]
    \centering
	\includegraphics[width=0.89\textwidth]{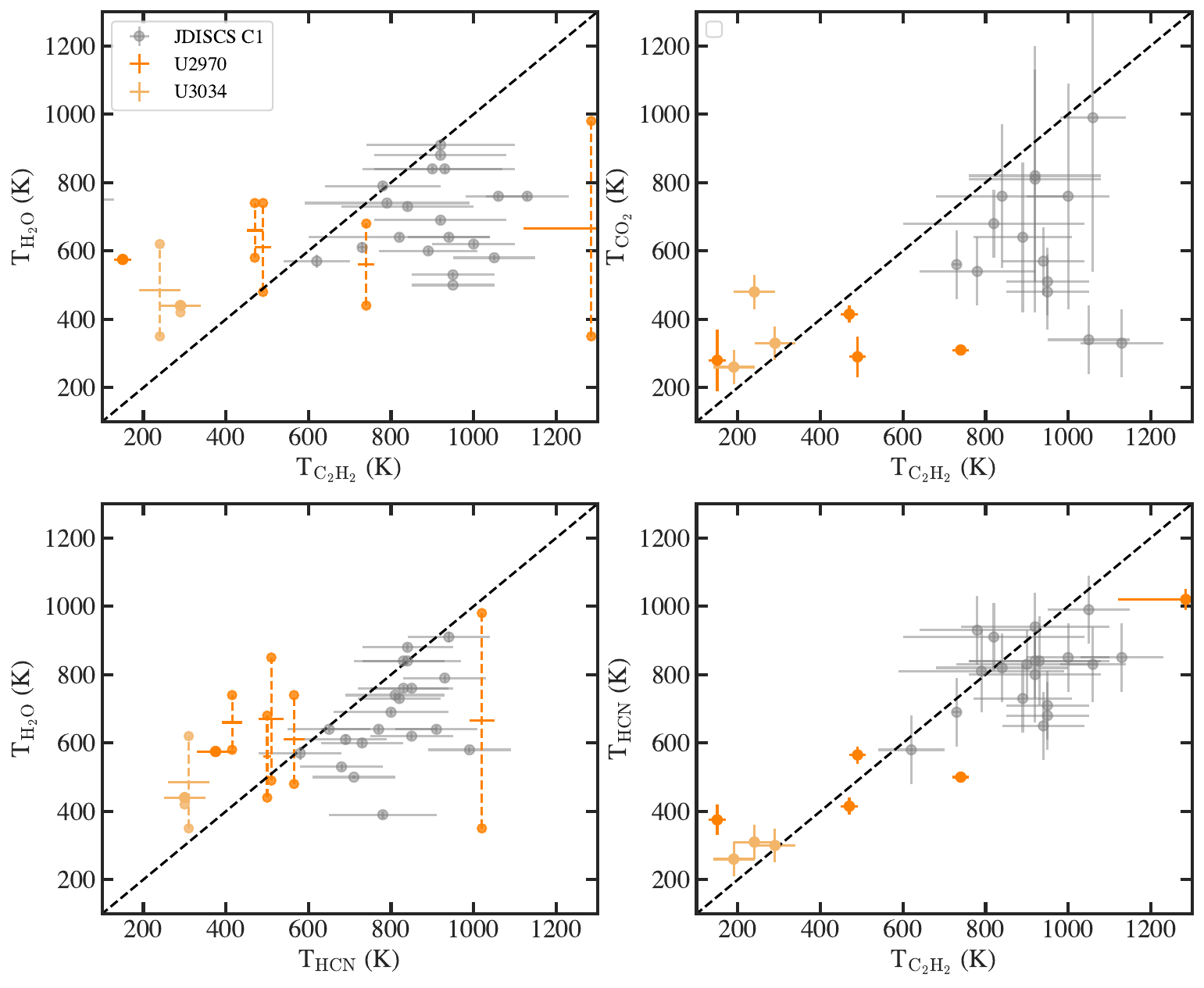}
    \caption{Temperature comparisons between molecules for each sample following the ratios plotted in Fig~\ref{fig:RatioMmol_vs_F89}. Except MR2, all USco disks show lower temperatures for each molecule compared to the young sample. 
    }
    \label{app:fig:T:comparison}
\end{figure}

Here, we provide a comparison of temperatures of the best fit models for the main molecular emissions among different samples, with the values of JDISCS C1 from \citet{Arulanantham25} and U3034 from Raul et al. in prep. For U2970 and U3034, if there are both hot and warm components of water, we take the hot and warm components as the upper and lower limits for the average temperature of the water emitting at $\sim10-19\,\mu$m.
We plot the comparison in Fig~\ref{app:fig:T:comparison} with corresponding molecules present in Fig.~\ref{fig:RatioMmol_vs_F89}. The temperatures of different molecules are generally consistent with each other (differences lower than two times the uncertainty). The temperatures in the older USco sample are consistently lower than the young JDISCS C1 sample. 

\begin{figure}[htb!]
    \centering
	\includegraphics[width=0.89\textwidth]{ 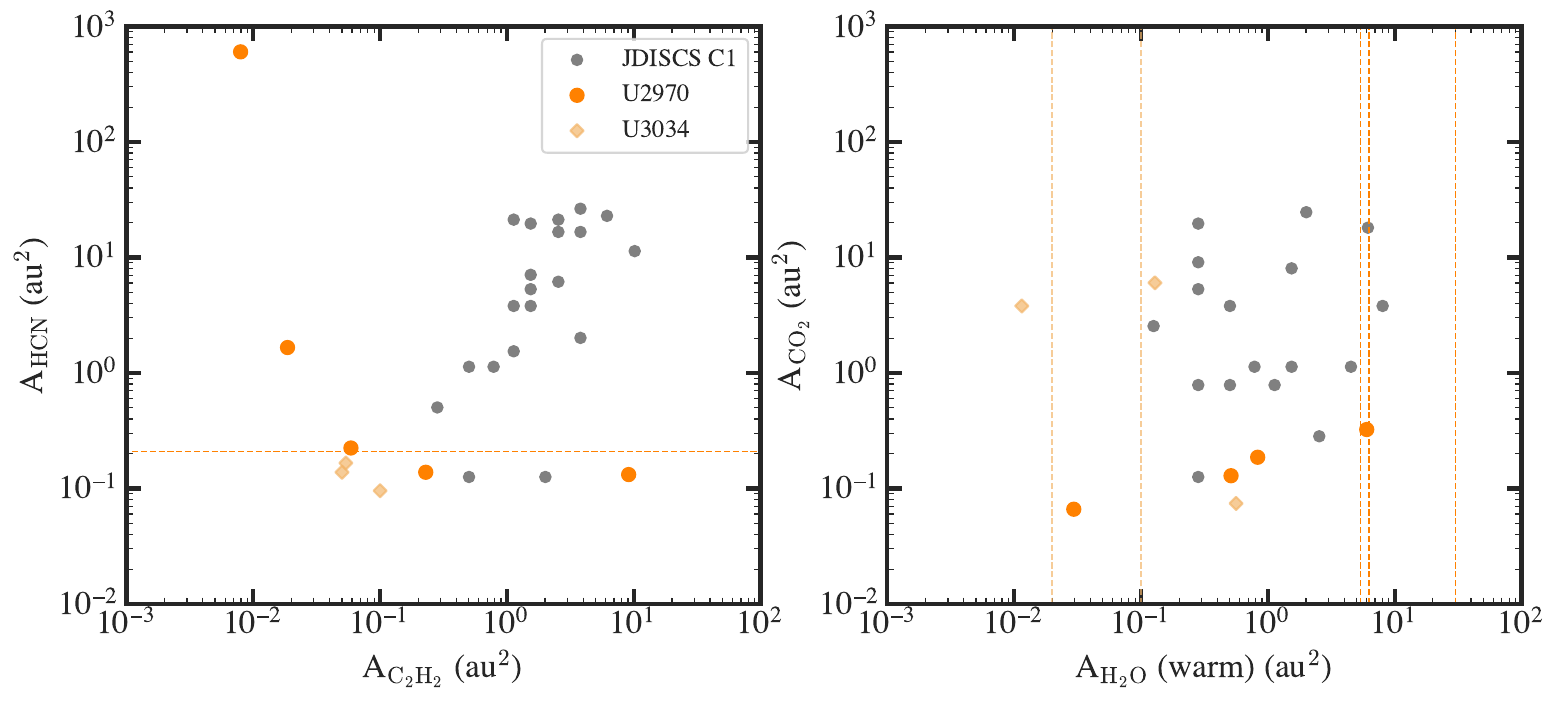}
    \caption{Comparison of best-fit emitting area for each of the molecules among samples. Dashed lines indicate the emitting area of one molecule if the other is not detected. For the organics, emitting areas are highly uncertain especially for the young JDISCS C1 sample where some of them are optically thin \citep[e.g.,][]{Arulanantham25}. For water, no significant differences can be seen between the young and older samples. 
    }
    \label{app:fig:R:comparison}
\end{figure}

The molecular emitting areas are less well constraint than the temperature and are degenerate with the column density. In addition, the emitting areas of C$_2$H$_2$ and HCN are degenerate with each other, because the emissions of these two species overlap in wavelengths and are fitted together.
Here we plot the fitted emitting areas of different samples in Fig~\ref{app:fig:R:comparison}, the degeneracy is clear and there is no significant difference between the old and young samples. 

\section{Optical depth of major molecular lines} \label{app:sec:tau}
In the LTE slab models, the optical depths of emission lines are crucial for the fitting and interpretation. Here, following the set-up of the models, we use the following equation to derive the optical depths at each line center \citep[e.g.,][]{Banzatti12,Tabone23} 
\begin{equation}
    \tau_{0} = \frac{\sqrt{\ln(2)}}{4\pi\sqrt{\pi}} \frac{A_{ul}N_{\rm mol} c^3}{\Delta v \nu_{ul}^3} (x_l\frac{g_u}{g_l}-x_u)
\end{equation}
Here, $x_i=g_i \exp(-E_i/kT_{\rm ex})/Q(T_{\rm ex})$. The $A_{ul}$, $g_i$, and $Q$, are the Einstein-A coefficient, statistical weight, and partition sum, respectively, which can be obtained for each single line from HITRAN database \citep{HITRAN22}, with $u$ and $l$ denoting the upper and lower energy level of each transition \citep[see][for more details]{Banzatti12}. $\Delta v$, $N_{\rm mol}$ and $T_{\rm ex}$ are the line width, column density and excitation temperatures from the models. 

For the main molecules (H$_2$O, C$_2$H$_2$, HCN and CO$_2$), and for the three representative temperatures (200\,K, 500\,K and 800\,K), we derive the column density where the optical depth at the main line center is one (see Table~\ref{app:tab:tau}). For water, we show the four single lines with hot, warm, and cool upper energy level \citep[][]{Banzatti25}. For C-bearing molecules, we show the strongest Q-branch line. The line width is assumed to be purely thermal with $\Delta v= \sqrt{\frac{k_BT}{M}}$. 
We note that the optical depth $\tau$ is proportional to the column density $N$, thus the corresponding $\tau$ for each of our models can be estimated directly. 

Based on these values, most of the water lines and Q-branches of the C-bearing molecules in the best-fit models (shown in Table~\ref{tab:fitting}) are optically thick. Generally, hot and warm water lines will become optically thin at $\log_{10}N<16$\,cm$^{-2}$, and lines from C-bearing molecules will become optically thin at $\log_{10}N\lesssim15$\,cm$^{-2}$.

\begin{deluxetable}{cc|ccc}
\tablecaption{Column density where the optical depth at the line center is one. 
}\label{app:tab:tau}
\tablewidth{0.49\textwidth}
\tablehead{Species & $\lambda$ ($\mu$m) & 800\,K & 500\,K & 200\,K \\
& & \multicolumn{3}{c}{$N_{\rm \tau=1}$}
}
\startdata
\multirow{4}{*}{H$_2$O}& 17.32 & 6.6$\times10^{16}$ & 9.8$\times10^{17}$ & 8.2$\times10^{23}$\\
 & 17.50 & 3.7$\times10^{16}$ & 8.1$\times10^{16}$ & 5.1$\times10^{19}$ \\
 & 23.817 & 1.7$\times10^{16}$ & 9.1$\times10^{15}$ & 1.3$\times10^{16}$ \\
 & 23.895 & 1.2$\times10^{16}$ & 7.4$\times10^{15}$ & 1.8$\times10^{16}$ \\
C$_2$H$_2$& 13.7 & 8.6$\times10^{15}$ & 2.0$\times10^{15}$ & 3.8$\times10^{14}$  \\
HCN& 14.02 & 1.5$\times10^{16}$ & 5.2$\times10^{15}$ & 1.4$\times10^{15}$ \\
CO$_2$& 14.97 & 1.45$\times10^{16}$ & 4.6$\times10^{15}$ & 1.1$\times10^{15}$ \\
\enddata
\tablecomments{$N_{\rm \tau=1}$ for different emitting temperature ($T_{\rm ex}$) and assuming a purely thermal line width. The four lines for water are the hot, warm and cool lines shown in Table~\ref{tab:fluxes} and \citet{Banzatti25}, and the values for the C-bearing molecules are the strongest Q-branch lines that dominate the flux. }
\end{deluxetable}

\bibliography{ref}{}
\bibliographystyle{aasjournal}



\end{document}